\documentclass[12pt]{article}
\usepackage[T1]{fontenc}
\usepackage[latin1]{inputenc}
\usepackage{geometry}
\usepackage{float}
\usepackage{mathrsfs}
\usepackage{amsmath}
\usepackage{amssymb}
\usepackage{graphicx}
\usepackage{setspace}
\usepackage{subfig}
\usepackage{hyperref}
\hypersetup{
    colorlinks=true,
    linkcolor=blue,
    filecolor=magenta,      
    urlcolor=cyan,
    citecolor = red,
}
\usepackage{gensymb}

\makeatletter

\floatstyle{ruled}
\newfloat{algorithm}{tbp}{loa}
\usepackage{algorithm}
\usepackage{algpseudocode}
\usepackage{caption}

\usepackage{bbm}

\usepackage{amsthm}

\usepackage{mathrsfs}

\usepackage{amsfonts}

\usepackage{epsfig}

\usepackage{bm}

\usepackage{mathrsfs}

\usepackage{enumerate}

\@ifundefined{definecolor}{\@ifundefined{definecolor}
 {\@ifundefined{definecolor}
 {\usepackage{color}}{}
}{}
}{}

\newtheorem{exam}{Example}[section]
\newtheorem{rem}{Remark}[section]

\providecommand{\algorithmname}{Algorithm}
\floatname{algorithm}{\protect\algorithmname}

\newcounter{hypA}

\newcounter{hypB}

\newcounter{hypD}

\makeatother

\usepackage{booktabs}
\usepackage{multirow}
\usepackage[table,xcdraw]{xcolor}


\renewenvironment{abstract}
 {\small
  \begin{center}
  \bfseries \abstractname\vspace{-.5em}\vspace{0pt}
  \end{center}
  \list{}{%
    \setlength{\leftmargin}{5mm}% <---------- CHANGE HERE
    \setlength{\rightmargin}{\leftmargin}%
  }%
  \item\relax}
 {\endlist}

\usepackage{blkarray}

\begin{document}
\emergencystretch 3em
%+Title
\begin{center}

\begin{doublespace}
{\Large \textbf{Sequential Markov Chain Monte Carlo for Lagrangian Data Assimilation with Applications to Unknown Data Locations}}
\end{doublespace}

\vspace{0.5cm}

BY HAMZA RUZAYQAT$^{1}$, ALEXANDROS BESKOS$^{2}$, DAN CRISAN$^{3}$, AJAY JASRA$^{1}$, \&  NIKOLAS KANTAS$^{3}$  \vspace{0.2cm}

{\footnotesize $^{1}$Applied Mathematics and Computational Science Program, Computer, Electrical and Mathematical Sciences and Engineering Division, King Abdullah University of Science and Technology, Thuwal, 23955-6900, KSA.}
\\
{\footnotesize $^{2}$Department of Statistical Science, University College London, London, WC1E 6BT, UK.}
\\
{\footnotesize $^{3}$Department of Mathematics, Imperial College London, London, SW7 2AZ, UK.}
\\
\end{center}

\begin{abstract}
We consider a class of high-dimensional spatial \textcolor{black}{filtering} problems, where the spatial locations of observations are unknown and driven by the partially observed hidden signal. This problem is exceptionally challenging as not only is high-dimensional, but the model for the signal yields longer-range time dependencies through the observation locations. Motivated by this model we revisit a lesser-known and \emph{provably convergent} computational methodology from \cite{berzuini, cent, martin} that uses sequential Markov Chain Monte Carlo (MCMC) chains. We extend this methodology for data filtering problems with unknown observation locations. We benchmark our algorithms on Linear Gaussian state space models against competing ensemble methods and demonstrate a significant improvement in both execution speed and accuracy. %\textcolor{black}{XXX FIX wording and dlogd}where the cost is of order $\mathcal{O}(dN + d_yN)$, $d$ is the state dimension, $d_y$ is the observations dimension and $N$ is the number of simulated samples. 
Finally, we implement a realistic case study on a high-dimensional rotating shallow water model (of about $10^4-10^5$ dimensions) with real and synthetic data. The data is provided by the National Oceanic and Atmospheric Administration (NOAA) and contains observations from ocean drifters in a domain of the Atlantic Ocean restricted to the longitude and latitude intervals $[-51\degree, -41\degree]$, $[17\degree, 27\degree]$ respectively.
\\ 
\noindent \textbf{MSC classes}:	62M20, 60G35, 60J20, 94A12, 93E11, 65C40

\noindent\textbf{Code available at:} \url{https://github.com/ruzayqat/filtering_unknown_data_locations}\\
\noindent\textbf{Corresponding author}: Hamza Ruzayqat. E-mail:
\href{mailto:ruzayqath@gmail.com}{ruzayqath@gmail.com}
\end{abstract}

%%%%%%%%%%%%%%%%%%%%%%%%%%%%%%%%%
\section{Introduction}

Consider a state-space model comprising of two elements, an unobserved continuous time stochastic process $(Z_t)_{t\geq 0}$ with $Z_t\in\mathbb{R}^d$
and a sequence of observations $(Y_{t_i})_{i\geq 1}$ taken at a sequence of known time instants $(t_i)_{i\geq 1}$. We will assume each $Y_{t_i}$ is a random variable depending on the evolution path of the unobserved process since the last time instant $(Z_t)_{t\in [t_{i-1},t_i]}$. $(Z_t)_{t\geq 0}$ and $(Y_{t_i})_{i\geq 1}$  are combined in a joint stochastic model
and the objective of \textcolor{black}{filtering}  is to estimate the unobserved state $Z_t$ given the observations up to that time
$(Y_{t_i})_{t_i\leq t}$. Such problems occur routinely in applications \textcolor{black}{in data assimilation} such as numerical weather prediction, oceanography, finance and engineering, {see \cite{carrassi, ghil, kalnay}} for example applications and \cite{bain,cappe} for book length introductions.

The problem of filtering is to approximate the conditional distribution of each $Z_{t_i}$ given
$(Y_{t_p})_{p\leq i}$ also known as the filtering distribution or simply the \emph{filter}. This is a challenging task
% even in the case that $Z_t\in\mathsf{Z}$, {where $\mathsf{Z}$ is a Borel subset of $\mathbb{R}^d$, and even when the state-space is small}. 
as in most cases of practical interest, with the exception of linear model observations and discrete-time, linear signal dynamics, the filter is not analytically available and hence one often has to resort to numerical approximations.
There are a plethora of techniques that have been presented in the literature, but perhaps the only two exact approaches (in the sense that the accuracy can be made arbitrarily high, subject to computational cost)  are based upon particle filters (PF) and Markov chain Monte Carlo (MCMC); see e.g.~\cite{bain,cappe,delm:13}. PFs and MCMC are simulation (or Monte Carlo) based approximations of the filter such that as the {the number of samples grows one} can recover the exact filter. PFs are specifically designed for filtering and have a fixed computational complexity per-time-step update. 

Whilst traditional iterative MCMC can be used for filtering at a fixed time, the cost grows at a linear order in the time step at each filtering update and is not often used for this task. Instead, if one wishes to explore this as an alternative direction, a more practical approach is to use sequential MCMC chains that target the filter at each time and also use the filter update equations.
Motivated by the challenges faced in high dimensional filtering and Lagrangian data assimilation we revisit a less popular computational methodology initially proposed in \cite{berzuini}. We note that this method is \emph{provably convergent} \cite{martin} and has been applied successfully in challenging filtering problems in the past such as for point process models \cite{cent,martin} and group tracking \cite{carmi,mihaylova}.

The problem of high-dimensional filtering is even more challenging. For instance {in numerical weather prediction models $d$ is in the order of millions}. Unfortunately simple or standard designs of PFs require an exponential cost in the dimension $d$ to achieve a certain precision and hence are impractical for very high dimensional applications. Several methods \cite{beskos1,beskos2,beskos3,kantas,kantas1,laggedPF} based upon a combination of sequential Monte Carlo (SMC) (e.g.~\cite{delm:06,delm:13}) and MCMC can be adopted. For well-designated classes of models, they have been shown to be both mathematically and empirically able to deal with high-dimensional filtering in a cost that is polynomial in the dimension of the problem. We note, however, that these methods are not a universal solution for the high-dimensional filtering problem due to the substantial computational cost involved. Several approximate (but biased) methods such as the ensemble-Kalman filter (EnKF), ensemble-transform Kalman filter (ETKF), error-subspace transform Kalman filter (ESTKF)  {and their localized versions} can be found in the literature, which despite being biased are the most popular and widely used methods for high-dimensional filtering, due to the very low computational cost relative to SMC-MCMC methods.

Motivated by problems in oceanography, we consider an even more complex high-dimensional filtering problem. In this case the observers travel on a spatial domain, such as a 2-dimensional grid, and their location is unknown and driven, deterministically, by the signal $(Z_t)_{t\geq 0}$ that is used to model the velocity field (among other variables) on the domain of interest. This is also known in the literature as Lagrangian Data Assimilation and this observation mechanism with unknown observer locations induces an extra-level of complexity versus the traditional \textcolor{black}{data assimilation} problem. The introduction of dependence of the spatial locations of observation on the signal yields a long-range (in time) dependence on the signal, that is not present in the classical filtering problem; the details are given in Section \ref{sec:model}. \textcolor{black}{Treating the position of observers as unknowns is a new challenge for filtering that has not been addressed in earlier works and} the ensemble methods mentioned above will not be accurate for this new and interesting problem. EnKF type methods are known to struggle with nonlinearities induced by Lagrangian observers even when the locations of the drifters are well known \cite{apte13} and the situation is much worse in the unknown location case. This was confirmed by extensive preliminary numerical simulations leading to this paper, which in addition showed that the computational cost required to implement SMC methods is very high due to the large number of tempering steps required. This motivated extending a sequential MCMC method developed for the filtering of point processes \cite{cent}, for this new type of filtering problem. The method of \cite{cent} has been shown in \cite{martin} to be a theoretically justified method for filtering (in the sense that the estimate will converge to the true filter as the number of samples grows to infinity) and seems to be particularly amenable for the filtering problem in high-dimensions, with a cost per update step depending \textcolor{black}{on the cost of evaluating the transition density of $Z_t$; e.g. for a homogeneous Gaussian transition density this is between $\mathcal{O}(d)$ and $\mathcal{O}(d^2)$ depending on its covariance matrix structure.  Also, in cases of regular grid and stationary Gaussian noise, costs can be $\mathcal{O}(d\log(d))$ even for dense covariance matrices by exploiting the Toeplitz structure of matrices involved  in the calculations and use fast Fourier transform (FFT), see e.g.~\cite{diet:93}.  
The known convergence guarantees cover the algorithm described later in the setting of known observation locations. 
In the case of unknown locations, the algorithm includes an extra approximation to address the dependence of the observer locations and the next target distribution to the random evolution of the state dynamics. Thus, for this case a complete analysis warrantees further work that is beyond the scope of the current paper. }

The contributions of this article are as follows:
\begin{enumerate}
\item{We formulate a new filtering problem with spatial Lagrangian observations at unknown locations. We develop, based upon \cite{cent}, a generic sequential MCMC method  that is effective for high-dimensional and nonlinear models. \textcolor{black}{We will often use the abbreviation SMCMC for the method in the sequel.}}
\item{We demonstrate the performance of this method in two ways. First we use tractable Linear Gaussian state space models and compare in terms of accuracy and execution speed with ensemble Kalman filter methods (EnKF, ETKF, ESTKF and  {Localized EnKF}) and show a significant improvement,  {especially when a higher accuracy is required}. Then we present a challenging realistic high dimensional data assimilation case study for which ensemble methods fail due to the nonlinearities present in the model and observation scheme. We use a rotating shallow water model and observations obtained both at known and unknown spatial locations. Our simulation scenarios are realistic and constructed using real data from NOAA.}
\end{enumerate}

This article is structured as follows. 
Section \ref{sec:model} describes the class of  models considered in this article.
Section \ref{sec:algo} presents the algorithms adopted for our problem of interest.
Section \ref{sec:numerics} demonstrates the methodology on several examples.  \textcolor{black}{Finally, in Section
\ref{sec:disc} we conclude the paper with a discussion.}

\section{Modelling Preliminaries}\label{sec:model}

%\subsection{State-Space Models and Filtering}\label{sec:ssm}

We consider an unknown continuous time stochastic process $(Z_t)_{t\geq 0}$, with $Z_t\in\mathsf{Z}\subseteq\mathbb{R}^d$, for which one has access to partial information via observations arriving at times, $\mathsf{T}:=\{t_1,t_2,\dots\}$, $t_0<t_1<t_2<\cdots$, $t_0=0$. At any given time $t_k\in \mathsf{T}$ we assume there are a fixed number of $d_y\in\mathbb{N}$ observations obtained from $N_d\in\mathbb{N}$ observers or drifters with the $j^{th}$ drifter's observation denoted by $y_{t_k}^j\in\mathsf{Y}$ and all observations collected {in the vector} ${y}_{t_k}\in\mathsf{Y}^{N_d}\subseteq \mathbb{R}^{d_y}$. 
These observations are {taken} at spatial locations $x_{t_1}^j,x_{t_2}^j,\dots$, $x_{t_k}^j\in\mathsf{X}\subseteq\mathbb{R}^{s}$, where $s$ is typically 2 or 3. The collection of spatial locations of all observers at an observation time $t_k\in\mathsf{T}$ is written as a vector ${x}_{t_k}\in\mathsf{X}^{N_d}$. 

%The particular structure of the model that is adopted is now described. 
We adopt a continuous time modelling approach for $Z_t$ motivated by applications such as atmosphere and ocean sciences. In these topics physical quantities such as wind or water velocity are modelled by continuous time space varying physical models comprising of systems of partial differential equations (PDEs). To allow for model uncertainty we need to incorporate stochastic dynamics for $Z_t$, for which the noise can be either added continuously (as in stochastic PDEs) or discretely in time (e.g. see Example \ref{exam:pde}). Our framework requires that at the discrete time instances in $\mathsf{T}$, $(Z_{t_k})_{k\geq 1}$ forms a discrete time Markov process with a known and well defined (positive) transition density. For $A\subseteq\mathsf{Z}$ we shall assume that the transition dynamics for $(Z_t)_{t\geq 0}$ can be obtained as
 
\begin{equation*}
\mathbb{P}(Z_{t_k}\in A|z_{t_{k-1}}) = \int_A f_k(z_{t_{k-1}},z_{t_k}) dz_{t_k}
\end{equation*}
where $\mathbb{P}$ denotes probability. {In the notation for} $f_k$ the subscript is included to account for possible time-inhomogeneous structure of $(Z_t)_{t\geq 0}$, or {dependence on the time increments} and $Z_0$ is taken as known. 
We will assume that, at the very least, $f_k(z_{t_{k-1}},z_{t_k})$ or a suitable approximation of it can be evaluated pointwise; examples include stochastic differential equations (SDEs) and their various time discretization approximations and similarly stochastic PDEs or PDEs with discrete time additive noise (see Example \ref{exam:pde} below). 

\begin{exam}[PDE with discrete time additive spatial noise]\label{exam:pde}
We present an example which will be considered often in the paper. We will consider $Z_t$ to be a vector containing hidden state variables at positions defined on a bounded domain with known boundary conditions.
Consider $\mathsf{Z}=\mathbb{R}^d$. Let $\Phi(z_s,s,t)$, $\Phi:\mathsf{Z}\times(\mathbb{R}^+)^2\rightarrow\mathsf{Z}$, be the solution of a PDE with initial condition $z_s$ run from time $s$ to $t$ with $0\leq s<t$. Then, an example of our model would be for $k\in\mathbb{N}$,
${Z_t} = \Phi(Z_{t_{k-1}},t_{k-1},t)$,  with ${t\in[t_{k-1},t_k)}$, and
\begin{equation*}
Z_{t_k}  = \Phi(Z_{t_{k-1}},t_{k-1},t_k)+ W_{t_k},
\end{equation*}
where $W_{t_k}\stackrel{\textrm{\emph{i.i.d.}}}{\sim}\mathcal{N}_d(0,\textcolor{black}{Q})$ is an i.i.d.~sequence of $d-$dimensional Gaussian random variables of zero mean and covariance matrix $\textcolor{black}{Q}$. In such scenarios, the process in continuous time is a PDE that is perturbed by noise at discrete times defined when the observations arrive and $f$ is a Gaussian density of mean $\Phi(z_{t_{k-1}},t_{k-1},t_k)$ and covariance matrix $\textcolor{black}{Q}$. Note that, in practice, one may have to replace $\Phi$ with a {numerical approximation of the solution of the PDE}.
\end{exam}

\subsection{The Standard Lagrangian Observation Model} \label{sec:standard_obs}
%MOVE: and for now we assume this is fixed and part of the data, omitting it from the notation. 
We will assume that each observation vector $Y_{t_k}$ depends only on $Z_{t_k}$ and there is a positive conditional likelihood density, i.e. for $k\in\mathbb{N}$, 
$B\subseteq \mathsf{Y}^{N_d}$,
%

%\begin{equation}
%\label{eq:obs_vanilla}
%\mathbb{P}\left({Y}_{t_k}\in B|(Z_t)_{t\geq 0}, (X_t)_{t\geq 0},({Y}_{t})_{t\in\mathsf{T}\setminus\{t_k\}}\right) = \int_B G(z_{t_k}, x_{t_k},{y}_{t_k}) d{y}_{t_k}=\int_B g_k(z_{t_k},{y}_{t_k}) d{y}_{t_k}.
%\end{equation}

\begin{equation}
\label{eq:obs_vanilla}
\mathbb{P}\left({Y}_{t_k}\in B|(Z_t)_{t\geq 0}, (X_t)_{t\geq 0},({Y}_{t})_{t\in\mathsf{T}\setminus\{t_k\}}\right) =\int_B g_k(z_{t_k},{y}_{t_k}) d{y}_{t_k}.
\end{equation}
\textcolor{black}{Note that in the standard model we discuss here, the position of the drifters is assumed known and for brevity their effect is captured simply by the subscript $k$ in $g_k$.} 
%Note in the standard case the observer locations, $x_{t_k}$, are known and part of the data, so in the notation we can use a time inhomogeneous conditional likelihood (and a subscript $k$) to denote this dependence.

\paragraph{Filtering and Smoothing}
Inference for the hidden state is performed using conditional distributions given the available data. One can either consider the whole path trajectory 
\begin{equation*}
\mathbb{P}(Z_{t_1},\ldots,Z_{t_k}|(X_{t_p})_{p\leq k},({Y}_{t_p})_{p\leq {k}}) \qquad \text{(smoothing)}
\end{equation*}
or just the marginal 
\begin{equation*}
\mathbb{P}(Z_{t_k}|(X_{t_p})_{p\leq k},({Y}_{t_p})_{p\leq {k}}) \qquad \text{(filtering)}.
\end{equation*}
%
%Often filtering is referred to as data assimilation and we will use both terms interchangeably. 
For $k\in\mathbb{N}$ we define the smoothing density
\begin{align}
\Pi_k(z_{t_{1}},\ldots, z_{t_k}) &\propto \prod_{p=1}^k f_p(z_{t_{p-1}},z_{t_p}) g_p(z_{t_p},{y}_{t_p}) \nonumber\\
&\propto f_k(z_{t_{k-1}},z_{t_k}) g_k(z_{t_k},{y}_{t_k}) \Pi_{k-1}(z_{t_{1}:t_{k-1}}). 
\label{eq:smooth_van_rec}
\end{align}
For $k\in\mathbb{N}$, we are interested in estimating expectations with respect to (w.r.t.) the filtering {distribution} (or simply the filter)
\begin{equation*}
\pi_k(\varphi):= \int_{\mathsf{Z}}\varphi(z_{t_k}) \pi_k(z_{t_k}) dz_{t_k},
\end{equation*}
where $\pi_k(z_{t_k})$ is the marginal in the $z_{t_k}$ co-ordinate of the {smoothing distribution} and $\varphi:\mathsf{Z}\rightarrow\mathbb{R}$ is $\pi_k-$integrable function. It easily follows that the filter density can be obtained recursively
\begin{equation}
\pi_k(z_{t_k}) \propto g_k(z_{t_k},{y}_{t_k})\int_{\mathsf{Z}}f_k(z_{t_{k-1}},z_{t_k})\pi_{k-1}(z_{t_{k-1}})dz_{t_{k-1}}.\label{eq:filter_recursion}    
\end{equation}
We note that the filtering problem is discrete time in nature due to the observations arriving at discrete times. One can still obtain $\mathbb{P}(Z_t|(X_{t_p})_{p\leq k},({Y}_{t_p})_{p\leq {k}})$ for $t\in (t_{k},t_{k+1})$ by integrating $\pi_{k-1}$ with the corresponding transition density (and applying Chapman-Kolmogorov equations).  

\subsection{State Space Models with Unknown Observer Locations}\label{sec:unknown_loc_model}

We proceed to extend the observation model to allow the spatial locations where observations are obtained to depend on the state process. For example, in Lagrangian Data Assimilation in oceanography the unknown ocean velocities will directly affect the drifter's motion. 
In particular, we now assume that for $j\in\{1,\dots,N_d\}$
\begin{equation}
dX_t^{j}  =  h(X_t^{j},Z_t)dt, \label{eq:obs_dynamics} 
\end{equation}
for some function $h:\mathsf{X}\times \mathsf{Z}\to \mathsf{X}$. We then modify the observation process, as originally considered in \eqref{eq:obs_vanilla}, for 
$k\in\mathbb{N}$, $B\subseteq \mathsf{Y}^{N_d}$
\begin{equation*}
\mathbb{P}\left({Y}_{t_k}\in B|(Z_t)_{t\geq 0}, ({Y}_{t})_{t\in\mathsf{T}\setminus\{t_k\}}\right) = 
\int_B G\left((z_{t_k},x_{t_k}),{y}_{t_k}\right) d{y}_{t_k},
\end{equation*}
where $G((z,x),\cdot)$ is a probability density on $\mathsf{Y}^{N_d}$; \textcolor{black}{we note that $G$ can depend upon the time parameter,  but we do not add this to the notation for simplicity and to distinguish with the known location case}. This observation model requires simulation of $x_{t_k}$ in parallel to $z_{t_k}$. 
Note that $(x_{t_k})_{ k \geq 1}$ is a deterministic function of $(z_t)_{t\leq t_k}$ and does not contain any additional information, but is required for the purpose of computing $G$ and needs to be propagated in the recursions for the dynamics. Compared to the model in \eqref{eq:obs_vanilla}, one has here that $G\left((z_{t_k},x_{t_k}),{y}_{t_k}\right)=g_k\left((z_t)_{t_{0}\leq t\leq t_k},y_{t_k}\right)$ instead.

\paragraph{Filtering and Smoothing} 

The filtering and smoothing recursions are analogous to the known observation location case and \eqref{eq:smooth_van_rec}-\eqref{eq:filter_recursion} in Section \ref{sec:standard_obs} where %$\mathbf{z}_{k}$ and $F_k,G$ replace $z_{t_k}$ and $f_k,g_k$.
$G$ replaces $g_k$.
% We note that a special case of this model, with no spatial locations $\mathbf{x}_{t_k}$ can be easily defined with a similar structure{. For instance, the signal can be a} discrete time Markov chain. Such an example is considered later on and we omit the details as they are essentially as above.
% The {smoothing density}, in this instance, can be written as
% %
% $$
% \pi_k(z_{t_1:t_k}) \propto \prod_{p=1}^k f_p(z_{t_{p-1}},z_{t_p}) G\left((z_{t_p},x_{t_p}),{y}_{t_p}\right).
% $$
% %
% Again, for $\pi_k-$integrable $\varphi:\mathsf{Z}\rightarrow\mathbb{R}$, $k\in\mathbb{N}$, we are interested in estimating expectations w.r.t.~the filter 
% %
% $$
% \pi_k(\varphi) := \int_{\mathsf{Z}}\varphi(z_{t_k}) \pi_k(z_{t_k}) dz_{t_k}
% $$
% %
% where $\pi_k(z_{t_k})$ is the marginal in the $z_{t_k}$ co-ordinate of the smoother.

\subsubsection{Discussion on the Choice of the Computational Methodology}
It should be clear from the previous discussion that since 
\begin{equation}\label{eq:x_sol_path}
{X_{t_k}^{j} = X_0^{j} + \int_{0}^{t_k} h(X_s^{j},z_s) ds}, \quad (j,k)\in \{1,\dots,N_d\} \times\mathbb{N},
\end{equation}
then 
%is, at a given time $t_k$ 
via the presence of $x_{t_k}$ the likelihood function $G((z_{t_k},x_{t_k}),\cdot)$  will depend upon the complete path realization of the hidden process $(Z_s)_{0\leq s \leq t_k}$ up to time $t_k$. This additional \emph{long-range dependence} is the barrier to using some of the existing particle filtering methods listed in the introduction. To explain this further we will first consider the case where one augments the state of the state space model with $(x_{t_k})_{ k \geq 1}$ and apply a standard PF, which is a sequential simulation method that propagate samples by combining importance sampling for \eqref{eq:smooth_van_rec} and resampling steps. Even for $d=1$ the standard PF, which is generally a reasonable method in that context, would be destined to fail, due to the well-known path degeneracy phenomenon \cite{kantas2}. This is a consequence of successive resampling steps that cause a lack of diversity in the samples representing the complete path $(Z_s)_{0\leq s \leq t_k}$ and approximating the corresponding smoothing density $\Pi_k$. On the other hand PFs can approximate very accurately the filter $\pi_k$ and this relies on the stability properties of the filter itself, which in turn requires reasonable mixing in the hidden state and a non degenerate likelihood function $G$ \cite{delm:13}. When augmenting hidden state dynamics with static parameters and implementing PFs, the deterministic components in the state vector cannot mix. Even when MCMC kernels are used as artificial dynamics they will depend on the complete path of the hidden state and path degeneracy will result into high Monte Carlo variance \cite{kantas2}. For this model with the deterministic dynamics $x_{t_k}$ an online PF implementation (with fixed computational cost per time) will suffer from both issues described: path degeneracy and lack of stochastic mixing. This is due to $(X_t)_{t\geq 0}$ being included in the information of $(Z_t)_{t\geq 0}$ and this justifies a separate treatment and algorithm design for this class of state space models. A more detailed theoretical study of filter properties for this model and its consequence for numerical methods is left for future work. \textcolor{black}{However, there are some exceptions where the joint  dynamics $(X_t,Z_t)_{t\geq 0}$ may display adequate mixing in continuous time as in some hypo-elliptic SDEs that can result in an ideal $f_k$ with sufficient mixing. Note that when time discretization is performed in these models the mixing in $(X_t)_{t\geq 0}$ will deteriorate or vanish unless recent advanced numerical schemes are used, e.g. as in \cite{monmarche, iguchi1, iguchi2}.}

%\begin{rem}
%There are some exceptions where the joint deterministic dynamics $(X_t,Z_t)_{t\geq 0}$ may display adequate mixing in continuous time as in some hypo-elliptic SDEs that can result in an ideal $f_k$ with sufficient mixing. Note that when time discretization is performed in these models the mixing in $(X_t)_{t\geq 0}$ will deteriorate or vanish unless recent advanced numerical schemes are used, e.g. as in \cite{monmarche}.
%\end{rem}

An alternative course of action is to aim for methods that aim to approximate $\pi_k$ without using importance sampling and resampling for $\Pi_k$ and \eqref{eq:smooth_van_rec}. The method in \cite{laggedPF}, which was designed for high-dimensional filtering, {has been} our first attempt to perform filtering for this model. The method transports particles from a variant of \eqref{eq:smooth_van_rec} that only considers the path between $t_{k-L+1}$ to $t_k$ for a small lag $L$ and thus bypasses the degeneracy issues by introducing a moderately low bias. % It is well suited to the problem, despite the dependence on the path of the signal described above. 
However, we found in extensive preliminary simulations, that the (computational) time to run such a methodology was significant and we aimed to lower the computational cost. Our investigation focused on efficiently approximating \eqref{eq:filter_recursion} directly via a sequence of MCMC chains initialized from a previously obtained approximation of $\pi_{k-1}$. %We will use a MCMC-type filtering algorithm, where 
At time $t_1$ the algorithm targets $\pi_1(z_{t_1})$ exactly by running an MCMC kernel with invariant distribution $\pi_1(z_{t_1})$ for $N$ steps, and at later times $t_k$, $k\geq 2$, it targets an approximation of the filter distribution obtained by replacing $\pi_{k-1}(z_{t_{k-1}})$ in \eqref{eq:filter_recursion} by an empirical density 
\begin{equation*}
S_{z,k-1}^N(z_{t_{k-1}}):=\frac{1}{N} \sum_{i=1}^N \textcolor{black}{\delta_{\left\{Z_{t_{k-1}}^{(i)}\right\}}}(z_{t_{k-1}}),
\end{equation*}
where \textcolor{black}{$\delta_{\left\{Z_{t_k}\right\}}(z)$ is the Dirac delta measure centered at $Z_{t_k}$}, and $Z_{t_{k-1}}^{(1)}, Z_{t_{k-1}}^{(2)}, \cdots, Z_{t_{k-1}}^{(N)}$ are the MCMC samples obtained in the preceding time step with the superscripts denoting MCMC iteration number. The method proved particularly effective and efficient in high dimensional problems and is trivially parallelizable.

\section{Sequential MCMC for Lagrangian Data Assimilation}\label{sec:algo}

\subsection{Standard State-Space Models}

We begin by detailing the \textcolor{black}{SMCMC} method in \cite{cent} in the case of the standard state-space model \textcolor{black}{(with the positions of the floaters assumed to be known)}. Our  setting differs from the one  for which the method was originally introduced. The approach is based upon the well-known prediction-updating structure that is given in \eqref{eq:filter_recursion}. 

The idea is to initiate the method with an MCMC algorithm
associated to an MCMC kernel of invariant measure

\begin{equation}
\pi_1(z_{t_1}) \propto f_1(z_0,z_{t_1})g_1(z_{t_1},{y}_{t_1}).
\label{eq:pi_1_known}
\end{equation}
\noindent 
\textcolor{black}{Note that we have assumed here for simplicity that $z_0$ is fixed and known.} 
There are many such kernels and in this article we focus on the random walk Metropolis (RWM) kernel exclusively. Such a kernel costs $\mathcal{O}(d)$ per step and requires one to be able to compute $f$ (or an approximation thereof).  \textcolor{black}{We note that RWM is a standard and popular generic choice of flexible kernel %when adequate MCMC performance is required, 
and in our numerical implementations shown in the sequel such a choice indeed provided an effective overall algorithm.} The MCMC kernel is run for $N$ steps producing $N$ samples and an $N-$sample approximation of $\pi_1(\varphi)$ given by
\begin{equation*}
\widehat{\pi}_1^N(\varphi) := \frac{1}{N}\sum_{i=1}^N \varphi(z_{t_1}^{(i)}),
\end{equation*}
where $z_{t_1}^{(i)}$ is the $i^{th}-$sample of the Markov chain.
At a subsequent time point $k\geq 2$, using \eqref{eq:filter_recursion}.
If one has access to an $N-$sample approximation of $\pi_{k-1}(z_{t_{k-1}})$, then instead of sampling from $\pi_k(z_{t_k})$ directly, which can be impossible, one can consider the approximate target density
\begin{equation}
\pi_k^N(z_{t_k}) \propto g_k(z_{t_k},{y}_{t_k}) \frac{1}{N} \sum_{i=1}^Nf_k(z_{t_{k-1}}^{(i)},z_{t_k}).
\label{eq:pi_k^N_known}
\end{equation}
One can then use any MCMC kernel (e.g.~RWM) with invariant measure $\pi_k^N$. Running the kernel for $N$ steps yields an $N-$sample approximation of $\pi_k(\varphi)$ as
\begin{equation*}
\widehat{\pi}_k^N(\varphi) := \frac{1}{N}\sum_{i=1}^N \varphi(z_{t_k}^{(i)}).
\end{equation*}
The method is summarized in Algorithm \ref{alg:1} (see Appendix \ref{sec:pseudocode} for a detailed pseudocode). We note that in our implementations, we initialize the MCMC chains (at time $k\geq 2$) by picking one of the samples from $\widehat{\pi}_{k-1}^N$ uniformly at random.

\begin{algorithm}
\begin{enumerate}
\item{Initialize: For any given initial distribution on $\mathsf{Z}$ run a MCMC kernel of invariant measure $\pi_1$ for $N$ steps. Keep in memory $\widehat{\pi}_1^N(\varphi)$. Set $k=2$.}
\item{Update: For any given initial distribution on $\mathsf{Z}$ run a MCMC kernel of invariant measure $\pi_k^N$ for $N$ steps. Keep in memory $\widehat{\pi}_k^N(\varphi)$. Set $k=k+1$. If $k=n+1$ go to step 3.~otherwise return to the start of step 2..}
\item{Return the approximations $\{\widehat{\pi}_k^N(\varphi)\}_{k\in\{1,\dots,n\}}$.}
\end{enumerate}
\caption{Sequential MCMC (SMCMC) Method for Filtering for $n$ time steps.}
\label{alg:1}
\end{algorithm}

The convergence of Algorithm \ref{alg:1} has been first discussed in \cite{martin}. \cite[Proposition 1]{martin} {gives} $\mathbb{L}_p-$bounds ($p\geq 1$) on the differences of $\widehat{\pi}_k^N(\varphi)-\pi_k(\varphi)$ of $\mathcal{O}(N^{-1/2})$, hence almost sure convergence of the estimators $\widehat{\pi}_k^N(\varphi)$ holds. It is important that the MCMC kernel possesses effective ergodicity properties and the better the kernel mixes, the better the approximations $\widehat{\pi}_k^N$ will be. %More recently \cite{finke} has presented additional results such as a central limit theorem and a discussion on when sequential MCMC can lead to lower asymptotic variance than traditional SMC methods. We believe that this is particularly relevant for high dimensional filtering problems and can be explored further in future work.

\textcolor{black}{The method as presented in Algorithm \ref{alg:1} has a  cost  of {$\mathcal{O}\left(\kappa(d)N^2+\kappa_y(d_y)N\right)$ per update step, where $\kappa(d)$ and $\kappa_y(d_y)$ are the costs for computing $f_k$ and $g_k$ respectively. The $\mathcal{O}(N^2)$ part of the cost can easily be reduced even to $\mathcal{O}(N)$ using subsampling of the samples used in \eqref{eq:pi_k^N_known},  with the overall cost then cut to $\mathcal{O}((\kappa(d)+\kappa_y(d_y))N)$}, using a simple auxiliary variable method on the sample indicator, which is what we will implement; {see Appendix \ref{sec:pseudocode} for details}. The cost due to dimensionality varies depending on the modeling, e.g.~in a Gaussian density as in Example \ref{exam:pde} calculating $f_k$ depends on the covariance matrix: $\kappa(d)=d$ if it is diagonal;  $\kappa(d)=d(d+1)/2$ for a general dense covariance matrix assuming a Cholesky decomposition can be performed offline and $Q$ is the same for every $k$; 
also, in the case of a regular grid and stationary Gaussian noise, one can exploit the Toeplitz structure of the matrices involved in calculations and use FFT, leading to a cost of $\kappa(d) =d\log(d)$. 
}

\subsection{State-Space Models with Unknown Observation Locations}

The approach detailed in the previous section is not straightforward to extend to the model with unknown observer locations in Section \ref{sec:unknown_loc_model}. The first issue is the integral in \eqref{eq:x_sol_path} 
% , that one has for $(k,j)\in\mathbb{N}\times\{1,\dots,d_y\}$
% %
% $$
% {X_{t_k}^j = X_{t_{k-1}}^j + \int_{t_{k-1}}^{t_k} h^j(X_s^j, Z_s)ds}
% $$
% %
% which 
is generally an intractable formula to compute. 
%and of course depends on the entire path of the process $(Z_s)_{s\in[0,t_k]}$.
One can replace the time integral by a simple Euler approximation with step size $\tau_{k}=(t_k-t_{k-1})/L$, $L\geq 2$: %and then one has a discretized location
\begin{equation}
{\tilde{X}_{t_k}^{j} = \tilde{X}_{t_{k-1}}^{j} +  \sum_{l=0}^{L-1}h(\tilde{X}_{t_{k-1}+l\tau_{k}}^{j}, Z_{t_{k-1}+l\tau_{k}})~\tau_{k}}\label{eq:x_tilde}    
\end{equation}
which still depends on a discrete path of the unobserved process.  \textcolor{black}{Note that more complex time discretization schemes can be used to improve numerical stability and accuracy relative to the simple Euler approximation.}

The next issue has to do with setting an appropriate target distribution for the MCMC sampler analogous to \eqref{eq:pi_k^N_known}. At time $t_{k-1}$, we have simulated samples $\{z_{t_{k-1}}^{(m)},x_{t_{k-1}}^{(m)}\}_{m=1}^{N}$ to plug in such an expression but the likelihood $G$ requires setting a single value for $x_{t_{k-1}}$ (as we can only have one target distribution for the MCMC chain). So we make one final approximation, and use a predicted value following:
\begin{equation}
\label{eq:loc_approx_rec}
\overline{x}_{t_k}^{j} = \overline{x}_{t_{k-1}}^{j} +  \mathbb{E}\Big[\sum_{l=0}^{L-1}h(\tilde{X}^{j}_{t_{k-1}+l\tau_{k}}, Z_{t_{k-1}+l\tau_{k}})\tau_{k}\Big|({Y}_{t_p})_{p\leq {k-1}},  Z_{t_{k-1}}= z_{t_{k-1}}, \tilde{X}^{j}_{t_{k-1}}=\overline{x}_{t_{k-1}}^{j} \Big]
\end{equation}
with the expectation taken w.r.t. $\big(\tilde{X}^{j}_{t_{k-1}+(l-1)\tau_k}, Z_{t_{k-1}+(l-1)\tau_k}\big)_{1<l\leq L}$ and can be approximated using plain Monte Carlo and propagating the dynamics of $Z_t$ jointly with \eqref{eq:x_tilde}.
%The definition of the terms $\overline{x}_{t_{0}}^j,  \overline{x}_{t_{1}}^j, \ldots$ is provided below.  
This will provide an approximation of the spatial locations in this manner for the next step without the need of elaborate simulation methods (like MCMC) that consider all of these discretization points. The main requirement is as before that the dynamics of $Z_t$ can be sampled, at least up-to an approximation. To illustrate this we present the procedure in Example \ref{exam:xbar} below for the case of Example \ref{exam:pde}.

We now proceed to present the analog of \eqref{eq:pi_k^N_known} appealing to the prediction-updating structure in \eqref{eq:filter_recursion}. The target distribution for the $k$-th MCMC procedure will be
\begin{equation*}
\overline{\pi}_k^N(z_{t_k}) \propto G\left((z_{t_k},{\overline{x}}_{t_k}),{y}_{t_k}\right) \frac{1}{N}\sum_{i=1}^N
f_k(z_{t_{k-1}}^{(i)},z_{t_k}).
\end{equation*}
Running the MCMC kernel for $N$ steps yields an $N-$sample approximation of $\overline{\pi}_k(\varphi)=\int_{\mathsf{Z}}\varphi(z_{t_k})\overline{\pi}_k(z_{t_k})dz_{t_k}$ as
\begin{equation*}
\widehat{\overline{\pi}}_k^N(\varphi) := \frac{1}{N}\sum_{i=1}^N \varphi(z_{t_k}^{(i)}).
\end{equation*}
The procedure is summarized in Algorithm \ref{alg:unknown}. %We note that \autoref{alg:unknown} can be adapted, for instance if $(Z_t)_{t\geq 0}$ is a continuous time process, with an appropriate approximation for the time discretization. Given the variety of processes and discretizations schemes we will not discuss this further in this article.

We remark that Algorithm \ref{alg:unknown} will have an intrinsic bias as it will (asymptotically) in $N$
approximate $\overline{\pi}_k$ and not the true filter. 
When the cost of the MCMC kernel is \textcolor{black}{$\mathcal{O}(\kappa(d))$} per iteration, then
 the cost of Algorithm \ref{alg:unknown} is \textcolor{black}{$\mathcal{O}(\kappa(d)N^2+LdN)$} per time step and the quadratic cost in $N$ can be removed as discussed previously to obtain a cost of \textcolor{black}{$\mathcal{O}((\kappa(d)+Ld)N)$}.

\begin{algorithm}
\begin{enumerate}
\item{Initialize: Compute $\overline{\mathsf{x}}_{t_1}$ using {the recursion formulae} \eqref{eq:loc_approx1}-\eqref{eq:loc_approx2} and for any given initial distribution on $\mathsf{Z}$ run a MCMC kernel of invariant measure $\overline{\pi}_1$ for $N$ steps. Keep in memory $\widehat{\overline{\pi}}_1^N(\varphi)$. Set $k=2$}
\item{Update: 
Compute $\overline{\mathsf{x}}_{t_k}$ using \eqref{eq:loc_approx3}-\eqref{eq:loc_approx4} and 
for any given initial distribution on $\mathsf{Z}$ run a MCMC kernel of invariant measure $\overline{\pi}_k^N$ for $N$ steps. Keep in memory $\widehat{\overline{\pi}}_k^N(\varphi)$. Set $k=k+1$. If $k=n+1$ go to step 3.~otherwise return to the start of step 2..}
\item{Return the approximations $\{\widehat{\overline{\pi}}_k^N(\varphi)\}_{k\in\{1,\dots,n\}}$.}
\end{enumerate}
\caption{SMCMC Method for Filtering with Unknown Locations for $n$ time steps.}
\label{alg:unknown}
\end{algorithm}

\begin{exam}[Example \ref{exam:pde} continued] \label{exam:xbar}
%The algorithm then proceeds as follows. 
To compute $\overline{x}_{t_1}^j$, this simply comprises of running the dynamics 
for $l\in\{0,\dots,L-2\}$
\begin{equation}\label{eq:loc_approx1}
Z_{(l+1)\tau_1}^{(r)} = \Phi(Z_{l\tau_1}^{(r)},l\tau_1,(l+1)\tau_1), \qquad r \in  \{1,\cdots,N\},
\end{equation}
and then (in parallel if possible) computing $\overline{x}_{t_1}^j$ using the recursion for $(l,j)\in\{0,\dots,L-1\}\times \{1\cdots,N_d\}$
\begin{equation}\label{eq:loc_approx2}
{{x}_{(l+1)\tau_1}^{j,(r)} = x_{l\tau_1}^{j,(r)}+ h\left(x_{l\tau_1}^{j,(r)}, Z_{l\tau_1}^{(r)}\right)~\tau_{1}},
\end{equation}
with initial condition $x_{t_{0}}^{j,(r)}=\overline{x}_{t_0}^{j}=x_{0}^{j}$ and setting 
\begin{equation}
\label{eq:loc_approx44}
{\overline{x}_{t_1}^{j} = \frac{1}{N}\sum_{r=1}^N x_{t_{1}}^{j,(r)}}.
\end{equation}
Then one runs a MCMC kernel {with} invariant measure:
\begin{equation*}
\overline{\pi}_1(z_{t_1}) \propto f_1(z_0,z_{t_1})G\left((z_{t_1},{\overline{x}}_{t_1}),{y}_{t_1}\right).
\end{equation*}
The MCMC kernel is run for $N$ steps and an $N-$sample proxy of 
$\overline{\pi}_1(\varphi)=\int_{\mathsf{Z}} \varphi(z_{t_1})\overline{\pi}_1(z_{t_1})dz_{t_1}$, assuming it is well-defined is
\begin{equation*}
\widehat{\overline{\pi}}_1^N(\varphi) := \frac{1}{N}\sum_{i=1}^N \varphi(z_{t_1}^{(i)}),
\end{equation*}
where $z_{t_1}^{(i)}$ is the $i^{th}-$sample of the MCMC chain.
At a subsequent time point $k\geq 2$, we now need to approximate the recursion \eqref{eq:loc_approx_rec}.
This can be achieved, by running for $(l,r)\in\{0,\dots,L-2\}\times\{1,\dots,N\}$
\begin{equation}\label{eq:loc_approx3}
Z_{t_{k-1}+(l+1)\tau_k}^{(r)} = \Phi(Z_{t_{k-1}+l\tau_k}^{(r)},t_{k-1}+l\tau_k,t_{k-1}+(l+1)\tau_k),
\end{equation}
and then for $(l,j,r)\in\{0,\dots,L-1\}\times \{1\cdots,N_d\}\times\{1,\dots,N\}$
\begin{equation}
\label{eq:loc_approx40}
{x_{t_{k-1}+(l+1)\tau_k}^{j,(r)} = x_{t_{k-1}+l\tau_k}^{j,(r)} +
h(x_{t_{k-1}+l\tau_k}^{j,(r)}, Z_{t_{k-1}+l\tau_k}^{(r)}) ~\tau_{k}},
\end{equation}
with initial condition $x_{t_{k-1}}^{j,(r)}=\overline{x}_{t_{k-1}}^{j}$, before finally obtaining 
\begin{equation}
\label{eq:loc_approx4}
{\overline{x}_{t_k}^{j} = \frac{1}{N}\sum_{r=1}^Nx_{t_{k}}^{j,(r)}}.
\end{equation}
\end{exam}

\section{Numerical Simulations}\label{sec:numerics}

We now illustrate the performance of Algorithms \ref{alg:1} and \ref{alg:unknown} in  {four} different cases:
\begin{enumerate}
    \item A tractable linear Gaussian model that is fully observed. The implementation of Algorithm~\ref{alg:1} has not been investigated for high-dimensional filtering problems and we will show both efficiency and accuracy compared to competing {ensemble methodologies}. In contrast to how practitioners use the latter, we use moderately high number of \textcolor{black}{ensemble sizes} ($500$-$10^4$) noting that for this model these methods are consistent and increasing \textcolor{black}{ensemble size} improves \emph{accuracy}.
    \item  {A tractable linear Gaussian model that is partially observed on a two-dimensional grid. In this model the dimension of the data is smaller than that of the unobserved hidden state. Here, we compare the SMCMC method against a local version of the ensemble Kalman filter where the physical domain is partitioned into subdomains and local updates are performed independently on each subdomain (in parallel).
    }
    \item A rotating shallow-water model observed at known locations. We use NOAA data to set the initial conditions and boundaries and then simulate $Z_t,x_t$ to provide observations. The point is to assess Algorithm \ref{alg:1} using synthetic drifter locations and observations, but we note the simulation scenario is set using real data from NOAA to make the case study as realistic as possible.
    \item A rotating shallow-water model observed with unknown locations. We use real data for observer positions and velocities and show that Algorithm \ref{alg:unknown} is effective at estimating both the unknown velocity fields and observer locations. 
\end{enumerate}

It is worth noting that in Cases 1.~{and 2.}~the true filter is known and is obtained through the Kalman filter (KF), however, in  {Cases 3.~and 4.}~the true filter is unknown, and therefore, in these two cases we compare our results to a reference that is chosen to be the hidden signal used to generate the observations in {Case 3.~and the prior distribution in Case 4.}~estimated using 50 different simulations of the shallow-water dynamics with noise as will be described below. \textcolor{black}{The convergence of the MCMC iterations is checked using standard diagnostics (trace and autocorrelation function plots, Gelman-Rubin diagnostic values), which for the sake of brevity are not reported here.}

\subsection{Linear-Gaussian Model: SMCMC vs Ensemble Methods}
\label{sec:linear_model1}
We begin with Algorithm \ref{alg:1} for a Linear and Gaussian model in discrete time (recall {that} this is related to the standard model described in Section \ref{sec:model}). Consider the model

\begin{equation}
\begin{split}
Z_{n+1} &= A Z_n + \sigma_z W_{n+1}, \quad W_{n+1} \stackrel{\textrm{i.i.d.}}{\sim} \mathcal{N}_d(0,I_d),  \quad n\in \{0,\cdots,T\},\\
 Y_m &= CZ_{mL} +\sigma_y V_m, \quad V_m \stackrel{\textrm{i.i.d.}}{\sim} \mathcal{N}_d(0,I_d), \quad m \in \{1, \cdots, T/L \},
 \end{split}
 \label{eq:linear_model}
\end{equation}
where $T\in \mathbb{N}$, $L\in \mathbb{N}$ is the time frequency at which the system is observed, i.e.~$t_j=jL$, 
$\mathsf{Z}:=\mathbb{R}^{d}$, 
$\mathsf{Y}=\mathbb{R}$, $d_y=d$, 
$A\in \mathbb{R}^{d \times d}$ is a square matrix \textcolor{black}{where} the maximum eigenvalue is less than or equal to one, $C\in \mathbb{R}^{d_y \times d}$ is defined as $C=[C_{i,j}]$ with
\begin{equation}
\label{eq:matrixC}
C_{i,j} = \left\{\begin{array}{lr}
1 & \text{if } j = i\,\hat{r}\\
0 & \text{otherwise}
\end{array} \right.,
\end{equation} 
where $\hat{r}$ is the spatial frequency at which the coordinates of $Z_{mL}$ are observed (e.g. if $\hat{r}=3$, we only observe the $3^{\textrm{rd}}$, $6^{\textrm{th}}$, $\cdots$ coordinates of $Z_{mL}$.) 

We will compare Algorithm \ref{alg:1} with the mean of the KF, EnKF, ETKF and ESTKF.
See \cite{ensembles} for the pseudo-codes. For the EnKF we use the matrix inversion lemma (Sherman-Morrison-Woodbury formula) when $d_y$ is larger than \textcolor{black}{the  ensemble size} (see \cite{nino,mandel} for more details.) 
In the implementation of the KF, the computation of the Kalman gain involves inverting a matrix that is known as the pre-fit residual. It is important to note that for all the ensemble methods mentioned in this article, we \emph{do not} directly compute the inverse of a matrix if it is being multiplied from left or right by another matrix or vector. Rather we reformulate the problem as finding the solution(s) of a linear system(s). For example, if we have $X=A^{-1} B$ (or $X=B A^{-1}$), we solve for $X$ in $A X = B$ (or in $A^T X^T = B^T$.)

\subsubsection{Simulation Settings}
For the simulations we set $T=500$, $L=1$, $\hat{r}=1$ (the system is fully observed), $A=0.2{I_d}$, $Z_0^j\sim -0.45 \times \mathcal{U}_{[0,1]}$ (uniform distribution on $[0,1]$), for $j\in\{1,\cdots,d\}$, and $\sigma_z=\sigma_y=0.05$, then compare the algorithms for different values of the state dimension $d$. The absolute errors are defined as the absolute difference of the mean of the KF and the means of the other filters at every time step $n$ and every state variable. We record the machine time once we have $70\%$ of the errors below $\sigma_y/2$. {We note that there are other possible metrics that could be used to evaluate the performance of the methods, but the chosen metric here is to log the machine time when $70\%$ of the absolute errors fall below a threshold value that we take it to be $\sigma_y/2$ (other threshold values are possible).}

\subsubsection{Results}
Table \ref{tbl:ensembles_vs_smcmc} displays the numerical results of our implementation. The table shows the state dimensions, the percentage of the absolute errors below $\sigma_y/2$, \textcolor{black}{the size of ensemble/particles} (plus $N_{burn}$ for SMCMC (Algorithm \ref{alg:1})), the number of independent simulations that were run, the machine time and the ratio of the machine time of ensemble methods w.r.t.~SMCMC. 
\begin{table}[h!]
\caption{(Model in Section \ref{sec:linear_model1}) Comparison of SMCMC and ensemble methods for different state dimensions $d$.}
\label{tbl:ensembles_vs_smcmc}
\centering
\resizebox{1.\textwidth}{!}{%
\begin{tabular}{@{}|c|ccccc||ccccccccccllll@{}}
\cmidrule(r){1-16}
\cellcolor[HTML]{C0C0C0}Methods & \multicolumn{1}{c|}{\textbf{KF}}        & \multicolumn{1}{c|}{\textbf{EnKF}} & \multicolumn{1}{c|}{\textbf{ETKF}} & \multicolumn{1}{c|}{\textbf{ESTKF}} & \textbf{SMCMC} & \multicolumn{1}{c|}{\textbf{KF}}        & \multicolumn{1}{c|}{\textbf{EnKF}} & \multicolumn{1}{c|}{\textbf{ETKF}} & \multicolumn{1}{c|}{\textbf{ESTKF}} & \multicolumn{1}{c||}{\textbf{SMCMC}} & \multicolumn{1}{c|}{\textbf{KF}}        & \multicolumn{1}{c|}{\textbf{EnKF}} & \multicolumn{1}{c|}{\textbf{ETKF}} & \multicolumn{1}{c|}{\textbf{ESTKF}} & \multicolumn{1}{c|}{\textbf{SMCMC}} &                      &                      &                      &                      \\ \cmidrule(r){1-16}
$d$                             & \multicolumn{5}{c||}{\textbf{625}}                                                                                                                                        & \multicolumn{5}{c||}{\textbf{1250}}                                                                                                                                                            & \multicolumn{5}{c|}{\textbf{4000}}                                                                                                                                                            &                      &                      &                      &                      \\ \cmidrule(r){1-16}
\% of Errors $\leq 0.5\sigma_y$ & \multicolumn{1}{c|}{}                   & \multicolumn{1}{c|}{0.730}         & \multicolumn{1}{c|}{0.729}         & \multicolumn{1}{c|}{0.729}          & 0.729          & \multicolumn{1}{c|}{}                   & \multicolumn{1}{c|}{0.720}         & \multicolumn{1}{c|}{0.721}         & \multicolumn{1}{c|}{0.719}          & \multicolumn{1}{c||}{0.716}          & \multicolumn{1}{c|}{}                   & \multicolumn{1}{c|}{0.720}         & \multicolumn{1}{c|}{0.720}         & \multicolumn{1}{c|}{0.721}          & \multicolumn{1}{c|}{0.720}          &                      &                      &                      &                      \\ \cmidrule(r){1-1} \cmidrule(lr){3-6} \cmidrule(lr){8-11} \cmidrule(lr){13-16}
\# of Ensemble/Particles       & \multicolumn{1}{c|}{}                   & \multicolumn{1}{c|}{500}           & \multicolumn{1}{c|}{500}           & \multicolumn{1}{c|}{500}            & $500 + 280$    & \multicolumn{1}{c|}{}                   & \multicolumn{1}{c|}{960}           & \multicolumn{1}{c|}{960}           & \multicolumn{1}{c|}{960}            & \multicolumn{1}{c||}{\textcolor{black}{ $200+900$}}     & \multicolumn{1}{c|}{}                   & \multicolumn{1}{c|}{3000}          & \multicolumn{1}{c|}{3000}          & \multicolumn{1}{c|}{3000}           & \multicolumn{1}{c|}{\textcolor{black}{$650+3000$}}    &                      &                      &                      &                      \\ \cmidrule(r){1-1} \cmidrule(lr){3-6} \cmidrule(lr){8-11} \cmidrule(lr){13-16}
\# of Simulations               & \multicolumn{1}{c|}{\multirow{-3}{*}{}} & \multicolumn{1}{c|}{1}             & \multicolumn{1}{c|}{1}             & \multicolumn{1}{c|}{1}              & 26             & \multicolumn{1}{c|}{\multirow{-3}{*}{}} & \multicolumn{1}{c|}{1}             & \multicolumn{1}{c|}{1}             & \multicolumn{1}{c|}{1}              & \multicolumn{1}{c||}{26}             & \multicolumn{1}{c|}{\multirow{-3}{*}{}} & \multicolumn{1}{c|}{1}             & \multicolumn{1}{c|}{1}             & \multicolumn{1}{c|}{1}              & \multicolumn{1}{c|}{26}             &                      &                      &                      &                      \\ \cmidrule(r){1-16}
Simulation Time (sec)           & \multicolumn{1}{c|}{18.6}               & \multicolumn{1}{c|}{102.5}         & \multicolumn{1}{c|}{44.1}          & \multicolumn{1}{c|}{39.4}           & 26.4           & \multicolumn{1}{c|}{61.7}               & \multicolumn{1}{c|}{268.7}         & \multicolumn{1}{c|}{160.3}         & \multicolumn{1}{c|}{144.4}          & \multicolumn{1}{c||}{\textcolor{black}{75.9}}           & \multicolumn{1}{c|}{1208.3}             & \multicolumn{1}{c|}{1628.8}        & \multicolumn{1}{c|}{1930.0}        & \multicolumn{1}{c|}{2048.8}         & \multicolumn{1}{c|}{\textcolor{black}{512.8}}          &                      &                      &                      &                      \\ \cmidrule(r){1-16}
Time ratios                     & \multicolumn{1}{c|}{0.70}               & \multicolumn{1}{c|}{3.89}          & \multicolumn{1}{c|}{1.67}          & \multicolumn{1}{c|}{1.49}           &                & \multicolumn{1}{c|}{\textcolor{black}{0.81}} & \multicolumn{1}{c|}{\textcolor{black}{3.54}}          & \multicolumn{1}{c|}{\textcolor{black}{2.11}}          & \multicolumn{1}{c|}{\textcolor{black}{1.90}}           & \multicolumn{1}{c||}{}               & \multicolumn{1}{c|}{\textcolor{black}{2.36}}               & \multicolumn{1}{c|}{\textcolor{black}{3.18}}          & \multicolumn{1}{c|}{\textcolor{black}{3.76}}          & \multicolumn{1}{c|}{\textcolor{black}{4.00}}           & \multicolumn{1}{c|}{}               &                      &                      &                      &                      \\ \cmidrule(r){1-16}
\cellcolor[HTML]{C0C0C0}Methods & \multicolumn{1}{c|}{\textbf{KF}}        & \multicolumn{1}{c|}{\textbf{EnKF}} & \multicolumn{1}{c|}{\textbf{ETKF}} & \multicolumn{1}{c|}{\textbf{ESTKF}} & \textbf{SMCMC} & \multicolumn{1}{c|}{\textbf{KF}}        & \multicolumn{1}{c|}{\textbf{EnKF}} & \multicolumn{1}{c|}{\textbf{ETKF}} & \multicolumn{1}{c|}{\textbf{ESTKF}} & \multicolumn{1}{c||}{\textbf{SMCMC}} & \multicolumn{1}{c|}{\textbf{KF}}        & \multicolumn{1}{c|}{\textbf{EnKF}} & \multicolumn{1}{c|}{\textbf{ETKF}} & \multicolumn{1}{c|}{\textbf{ESTKF}} & \multicolumn{1}{c|}{\textbf{SMCMC}} &                      &                      &                      &                      \\ \cmidrule(r){1-16}
$d$                             & \multicolumn{5}{c||}{\textbf{6250}}                                                                                                                                       & \multicolumn{5}{c||}{\textbf{9000}}                                                                                                                                                            & \multicolumn{5}{c|}{\textbf{12500}}                                                                                                                                                           & \multicolumn{1}{c}{} & \multicolumn{1}{c}{} & \multicolumn{1}{c}{} & \multicolumn{1}{c}{} \\ \cmidrule(r){1-16}
\% of Errors $\leq 0.5\sigma_y$ & \multicolumn{1}{c|}{}                   & \multicolumn{1}{c|}{0.702}         & \multicolumn{1}{c|}{0.703}         & \multicolumn{1}{c|}{0.701}          & 0.71           & \multicolumn{1}{c|}{}                   & \multicolumn{1}{c|}{0.709}         & \multicolumn{1}{c|}{0.709}         & \multicolumn{1}{c|}{0.708}          & \multicolumn{1}{c||}{0.706}          & \multicolumn{1}{c|}{}                   & \multicolumn{1}{c|}{0.732}         & \multicolumn{1}{c|}{0.731}         & \multicolumn{1}{c|}{0.731}          & \multicolumn{1}{c|}{0.728}          &                      &                      &                      &                      \\ \cmidrule(r){1-1} \cmidrule(lr){3-6} \cmidrule(lr){8-11} \cmidrule(lr){13-16}
\# of Ensemble/Particles       & \multicolumn{1}{c|}{}                   & \multicolumn{1}{c|}{4400}          & \multicolumn{1}{c|}{4400}          & \multicolumn{1}{c|}{4400}           & \textcolor{black}{$800+4600$}   & \multicolumn{1}{c|}{}                   & \multicolumn{1}{c|}{6500}          & \multicolumn{1}{c|}{6500}          & \multicolumn{1}{c|}{6500}           & \multicolumn{1}{c||}{\textcolor{black}{$1000+7000$}}   & \multicolumn{1}{c|}{}                   & \multicolumn{1}{c|}{10000}         & \multicolumn{1}{c|}{10000}         & \multicolumn{1}{c|}{10000}          & \multicolumn{1}{c|}{\textcolor{black}{$1500+10000$}}   & \multicolumn{1}{c}{} &                      &                      &                      \\ \cmidrule(r){1-1} \cmidrule(lr){3-6} \cmidrule(lr){8-11} \cmidrule(lr){13-16}
\# of Simulations               & \multicolumn{1}{c|}{\multirow{-3}{*}{}} & \multicolumn{1}{c|}{1}             & \multicolumn{1}{c|}{1}             & \multicolumn{1}{c|}{1}              & 26             & \multicolumn{1}{c|}{\multirow{-3}{*}{}} & \multicolumn{1}{c|}{1}             & \multicolumn{1}{c|}{1}             & \multicolumn{1}{c|}{1}              & \multicolumn{1}{c||}{26}             & \multicolumn{1}{c|}{\multirow{-3}{*}{}} & \multicolumn{1}{c|}{1}             & \multicolumn{1}{c|}{1}             & \multicolumn{1}{c|}{1}              & \multicolumn{1}{c|}{26}             & \multicolumn{1}{c}{} &                      &                      &                      \\ \cmidrule(r){1-16}
Simulation Time (sec)           & \multicolumn{1}{c|}{3487.7}             & \multicolumn{1}{c|}{3920.6}        & \multicolumn{1}{c|}{4547.4}        & \multicolumn{1}{c|}{4992.7}         & \textcolor{black}{1260.1}         & \multicolumn{1}{c|}{8797.8}             & \multicolumn{1}{c|}{9368.8}        & \multicolumn{1}{c|}{11546.3}       & \multicolumn{1}{c|}{12114.2}        & \multicolumn{1}{c||}{\textcolor{black}{2443.2}}         & \multicolumn{1}{c|}{20774.2}            & \multicolumn{1}{c|}{24932.3}       & \multicolumn{1}{c|}{36133.7}       & \multicolumn{1}{c|}{38533.4}        & \multicolumn{1}{c|}{\textcolor{black}{4800.3}}         &                      &                      &                      &                      \\ \cmidrule(r){1-16}
Time ratios                     & \multicolumn{1}{c|}{\textcolor{black}{2.77}}               & \multicolumn{1}{c|}{\textcolor{black}{3.11}}          & \multicolumn{1}{c|}{\textcolor{black}{3.61}}          & \multicolumn{1}{c|}{\textcolor{black}{3.96}}           &                & \multicolumn{1}{c|}{\textcolor{black}{3.6}}               & \multicolumn{1}{c|}{\textcolor{black}{3.83}}          & \multicolumn{1}{c|}{\textcolor{black}{4.73}}          & \multicolumn{1}{c|}{\textcolor{black}{4.96}}           & \multicolumn{1}{c||}{}               & \multicolumn{1}{c|}{\textcolor{black}{4.33}}               & \multicolumn{1}{c|}{\textcolor{black}{5.19}}          & \multicolumn{1}{c|}{\textcolor{black}{7.53}}          & \multicolumn{1}{c|}{\textcolor{black}{8.03}}           & \multicolumn{1}{c|}{}               &                      &                      &                      &                      \\ \cmidrule(r){1-16}
\cellcolor[HTML]{C0C0C0}Methods & \multicolumn{1}{c|}{\textbf{KF}}        & \multicolumn{1}{c|}{\textbf{EnKF}} & \multicolumn{1}{c|}{\textbf{ETKF}} & \multicolumn{1}{c|}{\textbf{ESTKF}} & \textbf{SMCMC} & \multicolumn{1}{l}{}                    & \multicolumn{1}{l}{}               & \multicolumn{1}{l}{}               & \multicolumn{1}{l}{}                & \multicolumn{1}{l}{}                & \multicolumn{1}{l}{}                    & \multicolumn{1}{l}{}               & \multicolumn{1}{l}{}               & \multicolumn{1}{l}{}                & \multicolumn{1}{l}{}                &                      &                      &                      &                      \\ \cmidrule(r){1-6}
$d$                             & \multicolumn{5}{c||}{\textbf{16000}}                                                                                                                                      &                                         &                                    &                                    &                                     & \multicolumn{1}{l}{}                & \multicolumn{1}{l}{}                    & \multicolumn{1}{l}{}               & \multicolumn{1}{l}{}               & \multicolumn{1}{l}{}                & \multicolumn{1}{l}{}                &                      &                      &                      &                      \\ \cmidrule(r){1-6}
\% of Errors $\leq 0.5\sigma_y$ & \multicolumn{1}{c|}{}                   & \multicolumn{1}{c|}{0.726}         & \multicolumn{1}{c|}{0.725}         & \multicolumn{1}{c|}{0.726}          & 0.721          & \multicolumn{1}{l}{}                    & \multicolumn{1}{l}{}               & \multicolumn{1}{l}{}               & \multicolumn{1}{l}{}                & \multicolumn{1}{l}{}                & \multicolumn{1}{l}{}                    & \multicolumn{1}{l}{}               & \multicolumn{1}{l}{}               & \multicolumn{1}{l}{}                & \multicolumn{1}{l}{}                &                      &                      &                      &                      \\ \cmidrule(r){1-1} \cmidrule(lr){3-6}
\# of Ensemble/Particles       & \multicolumn{1}{c|}{}                   & \multicolumn{1}{c|}{16500}         & \multicolumn{1}{c|}{16500}         & \multicolumn{1}{c|}{16500}          & \textcolor{black}{$2000+13000$}   &                                         & \multicolumn{1}{l}{}               & \multicolumn{1}{l}{}               & \multicolumn{1}{l}{}                & \multicolumn{1}{l}{}                & \multicolumn{1}{l}{}                    & \multicolumn{1}{l}{}               & \multicolumn{1}{l}{}               & \multicolumn{1}{l}{}                & \multicolumn{1}{l}{}                &                      &                      &                      &                      \\ \cmidrule(r){1-1} \cmidrule(lr){3-6}
\# of Simulations               & \multicolumn{1}{c|}{\multirow{-3}{*}{}} & \multicolumn{1}{c|}{1}             & \multicolumn{1}{c|}{1}             & \multicolumn{1}{c|}{1}              & 26             &                                         & \multicolumn{1}{l}{}               & \multicolumn{1}{l}{}               & \multicolumn{1}{l}{}                & \multicolumn{1}{l}{}                & \multicolumn{1}{l}{}                    & \multicolumn{1}{l}{}               & \multicolumn{1}{l}{}               & \multicolumn{1}{l}{}                & \multicolumn{1}{l}{}                &                      &                      &                      &                      \\ \cmidrule(r){1-6}
Simulation Time (sec)           & \multicolumn{1}{c|}{43328.3}            & \multicolumn{1}{c|}{44263.8}       & \multicolumn{1}{c|}{69766.3}       & \multicolumn{1}{c|}{73085.9}        & \textcolor{black}{8960.1}        & \multicolumn{1}{l}{}                    & \multicolumn{1}{l}{}               & \multicolumn{1}{l}{}               & \multicolumn{1}{l}{}                & \multicolumn{1}{l}{}                & \multicolumn{1}{l}{}                    & \multicolumn{1}{l}{}               & \multicolumn{1}{l}{}               & \multicolumn{1}{l}{}                & \multicolumn{1}{l}{}                &                      &                      &                      &                      \\ \cmidrule(r){1-6}
Time ratios                     & \multicolumn{1}{c|}{\textcolor{black}{4.84}}               & \multicolumn{1}{c|}{\textcolor{black}{4.94}}          & \multicolumn{1}{c|}{\textcolor{black}{7.79}}          & \multicolumn{1}{c|}{\textcolor{black}{8.16}}           &                & \multicolumn{1}{l}{}                    & \multicolumn{1}{l}{}               & \multicolumn{1}{l}{}               & \multicolumn{1}{l}{}                & \multicolumn{1}{l}{}                & \multicolumn{1}{l}{}                    & \multicolumn{1}{l}{}               & \multicolumn{1}{l}{}               & \multicolumn{1}{l}{}                & \multicolumn{1}{l}{}                &                      &                      &                      &                      \\ \cmidrule(r){1-6}
\end{tabular}
}
\end{table}

\begin{rem}
Simulations were run on a Precision 7920 Tower Workstation with 52 cores and 512GB of memory. We note that in our method matrix multiplication is very limited and there is no matrix inversion. Each run of any of the ensemble methods is allowed to use all 52 cores for matrix multiplication through multi-threading, while each run of our method only uses one core, however, we run 26 independent simulations of SMCMC (for the model in Section \ref{sec:linear_model1}) in parallel and then average the results. 
\end{rem}

\begin{figure}[h!]
\begin{minipage}[b]{.485\textwidth}
\centering
\includegraphics[scale=0.36]{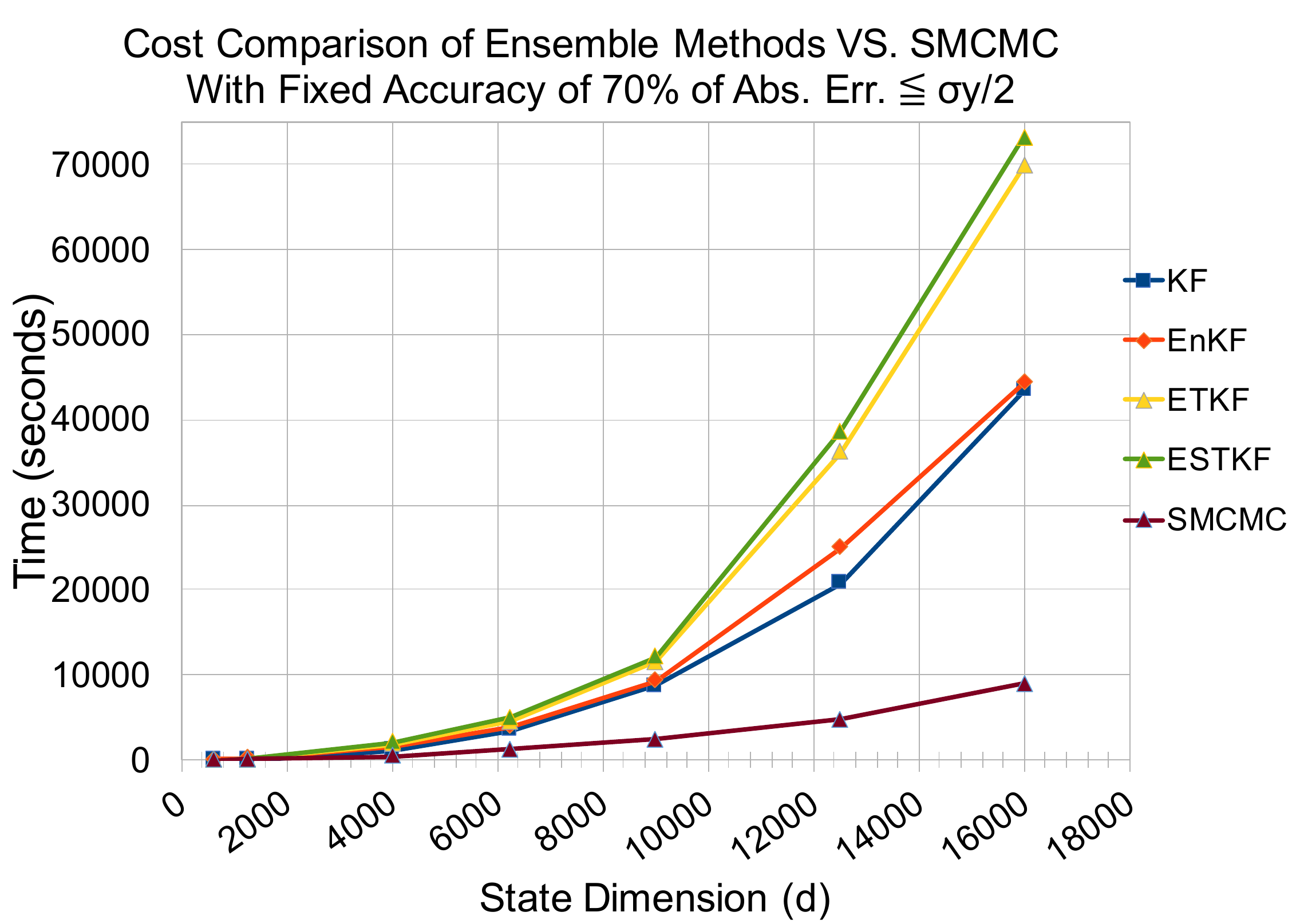}
\caption{(Model in Section \ref{sec:linear_model1}.) Comparison of machine times of SMCMC versus ensemble methods for different state dimensions $d$ at fixed accuracy. This is the total time spent on all simulations when the fraction of absolute errors smaller than $\sigma_y/2$ is approximately 70\%.}
\label{fig:ensembles_vs_smcmc}
\end{minipage}
\hfill
\begin{minipage}[b]{.485\textwidth}
\centering
\includegraphics[scale=0.36]{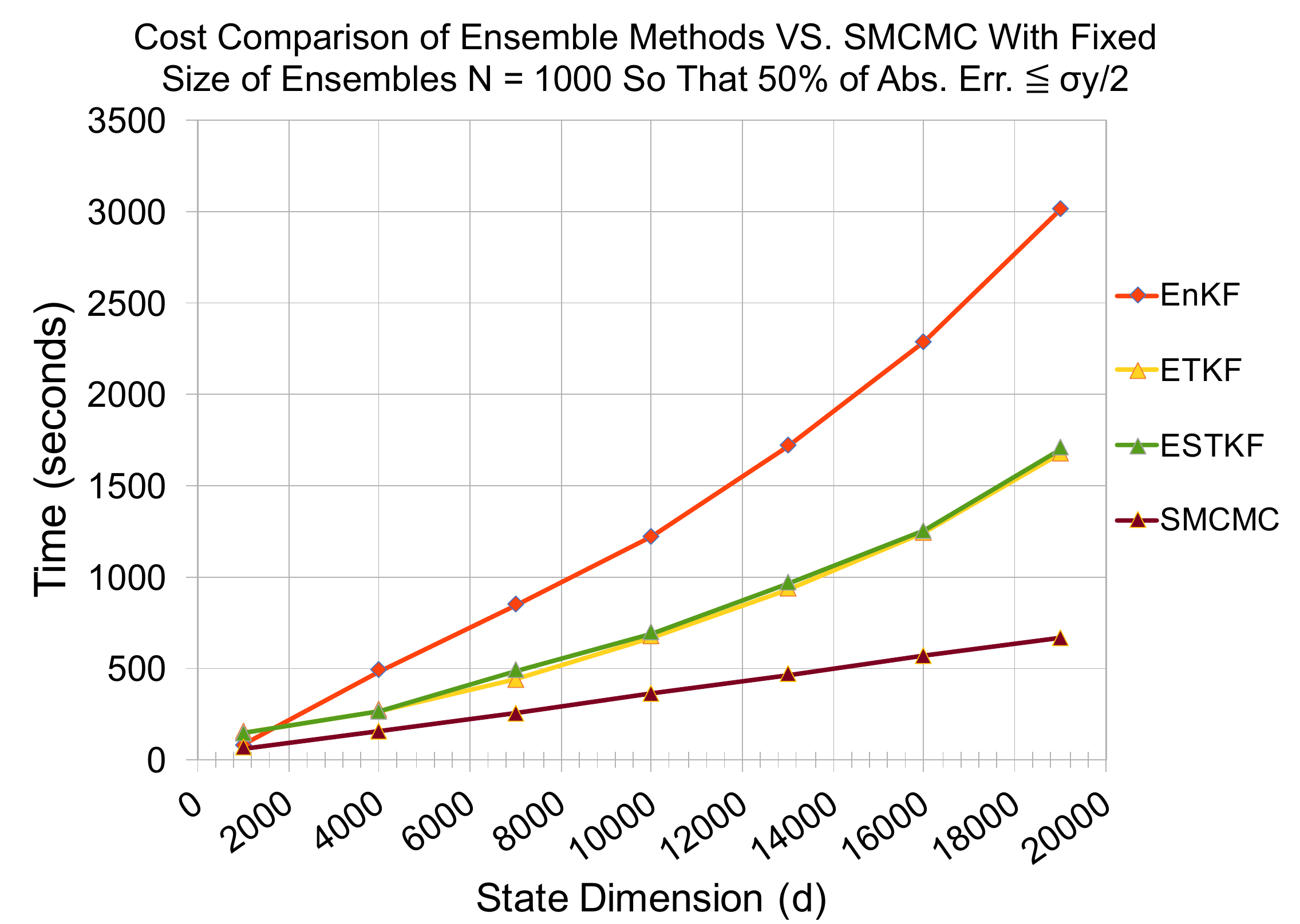}
\caption{(Model in Section \ref{sec:linear_model1}.) Comparison of machine times of SMCMC versus ensemble methods for different state dimensions $d$ when the \textcolor{black}{size of ensemble/particles} and accuracy are fixed. We set $N=1000$ such that \textcolor{black}{at least} 50\% of the absolute errors are less than $\sigma_y/2$. }
\label{fig:ensembles_vs_smcmc1}
\end{minipage}
\end{figure}

In Figure \ref{fig:ensembles_vs_smcmc}, we plot the machine times for each method. In Figure \ref{fig:ensembles_vs_smcmc1}, we fix the size of ensemble/particles to $N=1000$. We choose $N$ this way such that \textcolor{black}{at least} 50\% of the absolute errors \textcolor{black}{of each method} are less than $\sigma_y/2$. The plot shows that even when \textcolor{black}{the  ensemble size} is low and fixed the SMCMC method is still dominating especially as the state dimension grows to high values. The results indicate that Algorithm \ref{alg:1} is superior to several established data assimilation methods, particularly in high-dimensional scenarios. For example, when $d=16000$, achieving a success rate of 72\% in reducing errors below $\sigma_y/2$ using the EnKF is almost \textcolor{black}{five} times as expensive as using the SMCMC method. Additionally, the cost for achieving the same level of performance is \textcolor{black}{almost eight times greater for the ETKF and ESTKF methods}. These findings provide clear evidence for the superior performance of Algorithm \ref{alg:1} (when compared to non-localized ensemble methods) in high-dimensional data assimilation problems.

\subsection{Linear-Gaussian Model: SMCMC vs Local EnKF Method}
\label{sec:linear_model2}
 {Consider the same model as in \eqref{eq:linear_model} on a square grid such that $d=N_s^2$, $d_y = N_o^2$, for some $N_s, N_o\in \mathbb{N}$. In this example we compare the SMCMC method detailed in Algorithm \ref{alg:1} with a local version of the EnKF method. Typically, localization is implemented in two ways: explicitly, by considering only observations within a specific region surrounding the analysis location, or implicitly, by modifying the analysis state so that observations beyond a certain distance do not impact it. The application of localization is an ongoing area of research, and numerous variations of localization schemes have emerged in the past decade (see \cite{ensembles} and the references therein). The localization technique used here is an explicit technique and reffered to as the $R$-localization ($R$ refers to the observational error covariance matrix). The basic idea of $R$-localization is not to use all the grid points when updating the ensemble members but to perform a domain localization where the domain $D$ is partitioned into $\Gamma$ subdomains $\{D_i\}_{i=1}^{\Gamma}$, then, to only use the observations within a specified distance (the localization radius) around the local domain $D_i$. This defines a linear transformation $\widehat{D}_i$, which transforms the observation error covariance matrix, the global observation vector $Y_m$, and the global observation matrix $C$ to their local parts. Then, an observation localization is performed by modifying the observation error covariance matrix so that the inverse observation variance decreases to zero with the distance of an observation from an analysis grid point. The modification is done by dividing the diagonal elements of $R$ by some weights that are a function (e.g. Gaspari--Cohn function or an exponential function) of the distance of observations from the analysis grid point (for more details see \cite{ensembles}). The updates on the local domains can be done independently and therefore in parallel.}  \textcolor{black}{Note that the model does not require to enforce spatial smoothness/continuity on the hidden state (in constrast, e.g., to the Rotating Shallow-Water model used in the sequel), thus the localisation approach permits jumps between neighbouring subdomains reconstructed by ensemble members.}

\subsubsection{Simulation Settings}
 {Let $T=100$, $L=1$, $\hat{r}=4$ (that is, only one-fourth of the grid points are observed, i.e., the observed coordinates are 
$4, 8, 12, \cdots$). The matrix $C$ is the same as in \eqref{eq:matrixC}, $A = -0.95 I_d$, $\sigma_y=\sigma_z=0.1$, and we set $Z_0^j\sim -0.15 \times \mathcal{U}_{[0,1]}$ \textcolor{black}{for all $j\in \{1,\cdots,\lfloor d/3 \rfloor\}$ and $Z_0^j = 0$ for the rest of $j$'s}. First, we test all algorithms (SMCMC, EnKF, ETKF, ESTKF and local EnKF (LEnKF)) when $d=1600$ and $d_y=400$ and plot the errors in histograms. Then we change the values of the state dimension $d$ (and hence $d_y$) and compare the SMCMC algorithm against the local EnKF, with $R$-localization, in terms of the cost for fixed accuracy by recording the machine time needed by both algorithms once the percentage of absolute errors less than $\sigma_y/2$ is 70\%. We also compute the cost when the percentage of absolute errors less than $\sigma_y/2$ is 77\%. We note that Algorithm \ref{alg:1} is run $M$ times in parallel on a 52 cores machine. As for the local EnKF, the parallelization is done over the $\Gamma$ subdomains since the updates on the local domains are independent. We note that the number of subdomains $\Gamma$ is chosen between 15 and 55 such that it divides $d$.}

\subsubsection{Results}
\begin{figure}[ht]
\begin{minipage}[b]{.485\textwidth}
\centering
\includegraphics[scale=0.15]{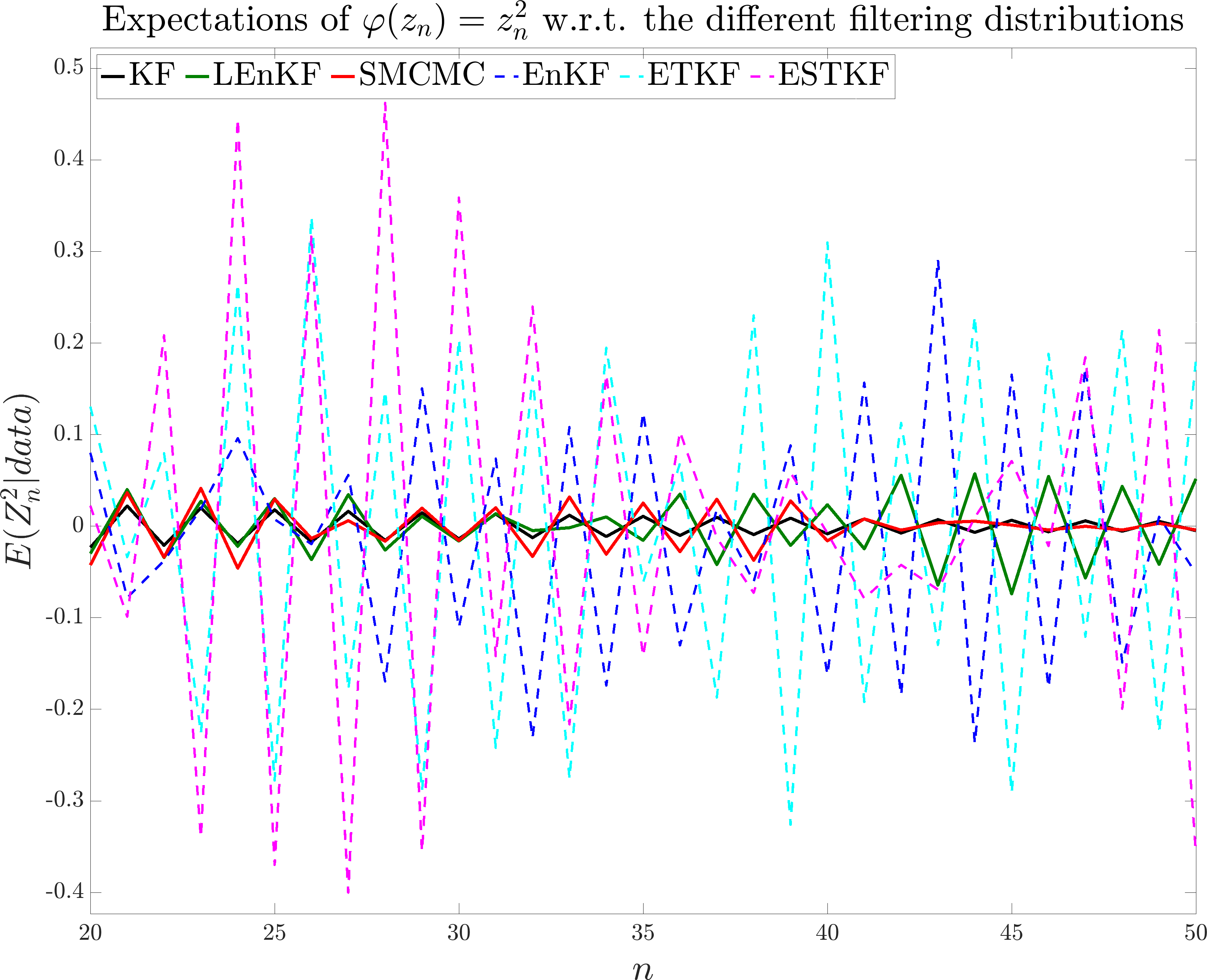}
\caption{ {(Model in Section \ref{sec:linear_model2}.) This illustrates the means of the KF (truth), EnKF, ETKF, ESTKF, LEnKF and SMCMC for the \emph{unobserved} state variable $Z_n^{(2)}$ for all $n \in \{20,\cdots,50\}$.}}
\label{fig:smcmc_vs_lenkf_coord2_all}
\end{minipage}
\hfill
\begin{minipage}[b]{.485\textwidth}
\centering
\includegraphics[scale=0.154]{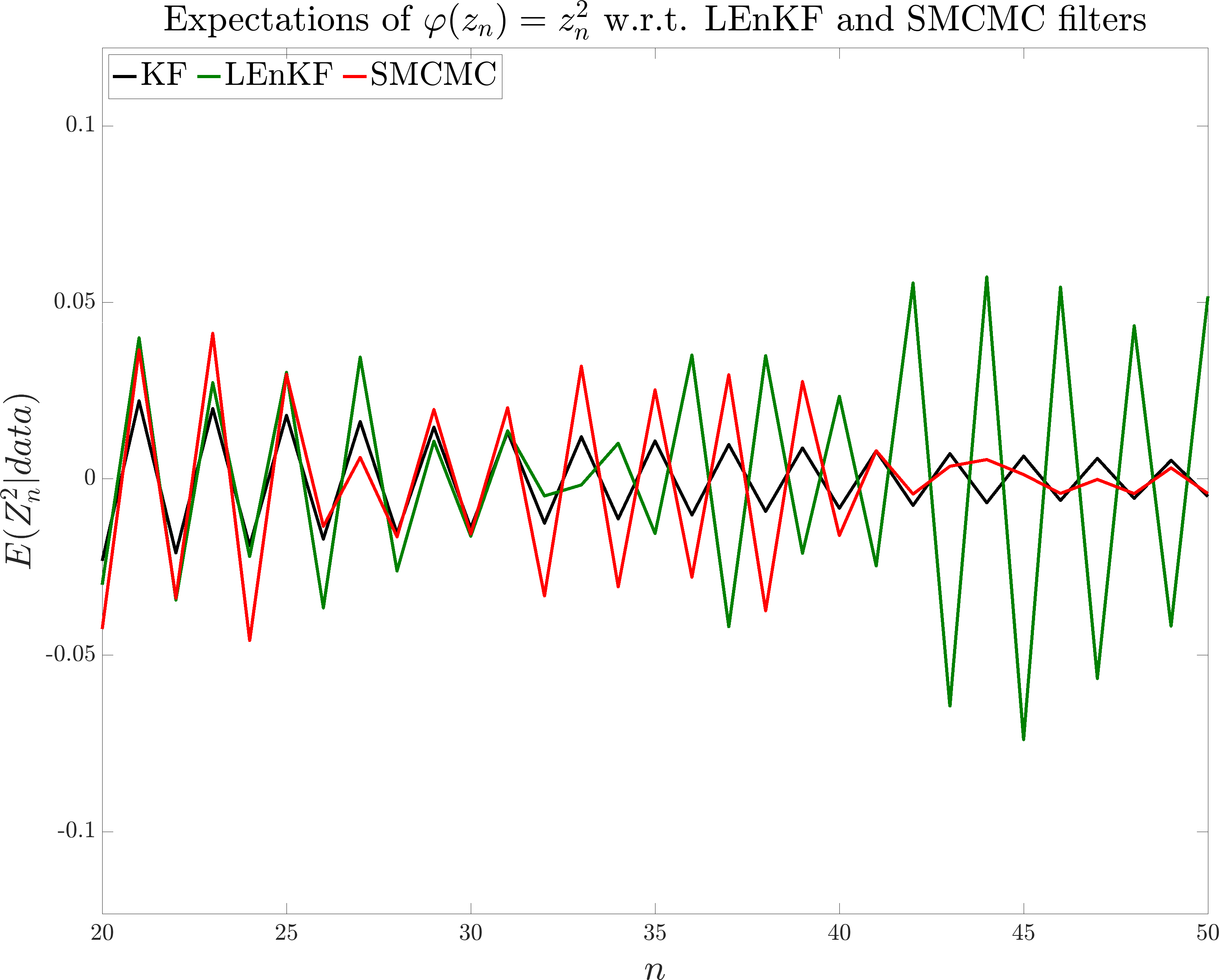}
\caption{ {(Model in Section \ref{sec:linear_model2}.) This illustrates the means of the KF (truth), LEnKF and SMCMC for the \emph{unobserved} state variable $Z_n^{(2)}$ for all $n \in \{20,\cdots,50\}$.}}
\label{fig:smcmc_vs_lenkf_coord2}
\end{minipage}
\end{figure}

\begin{figure}[ht]
\begin{minipage}[b]{.485\textwidth}
\centering
\includegraphics[scale=0.15]{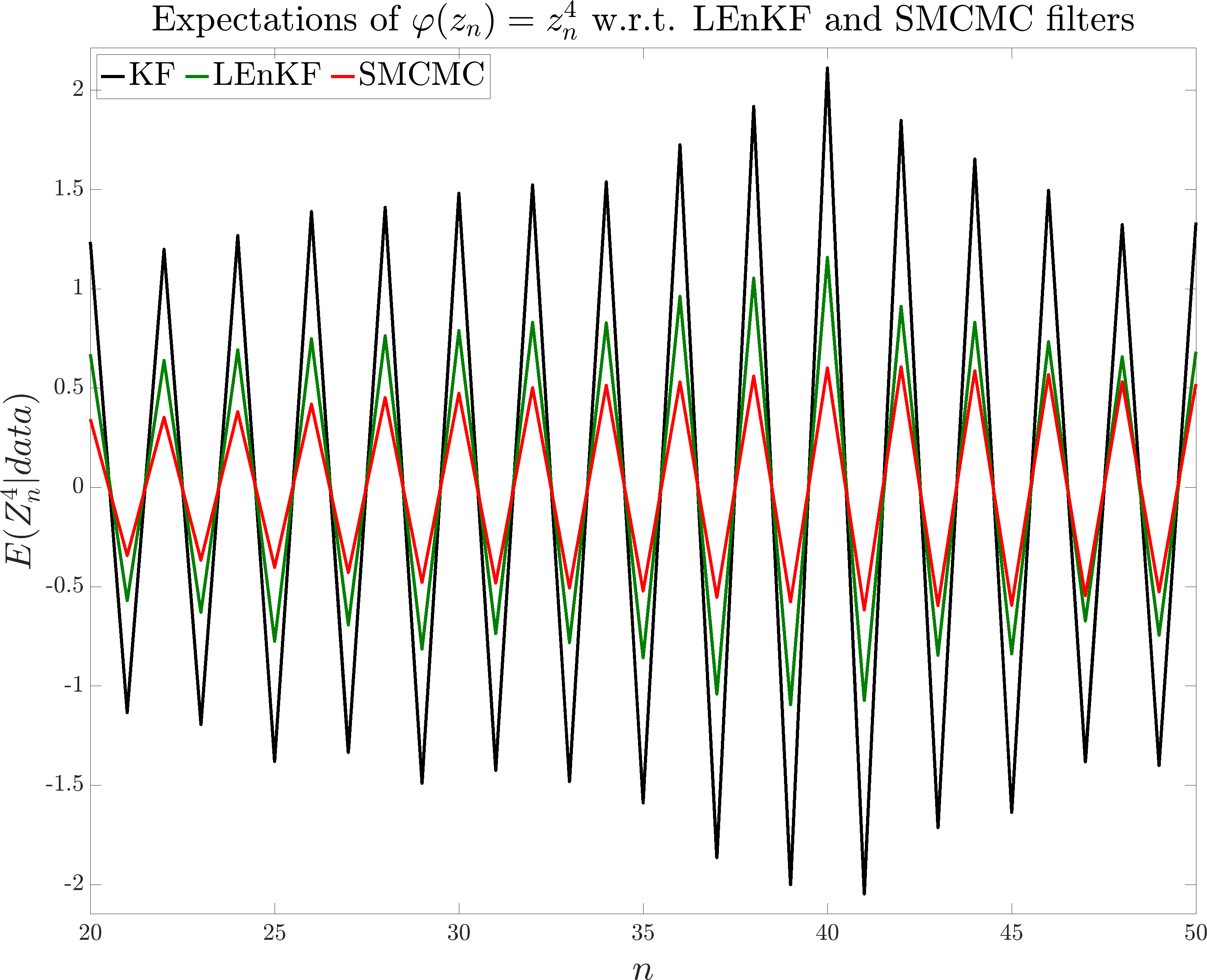}
\caption{ {(Model in Section \ref{sec:linear_model2}.) This illustrates the means of the KF (truth), LEnKF and SMCMC for the \emph{observed} state variable $Z_n^{(4)}$ for all $n \in \{20,\cdots,50\}$.}}
\label{fig:smcmc_vs_lenkf_coord4}
\end{minipage}
\hfill
\begin{minipage}[b]{.485\textwidth}
\centering
\includegraphics[scale=0.15]{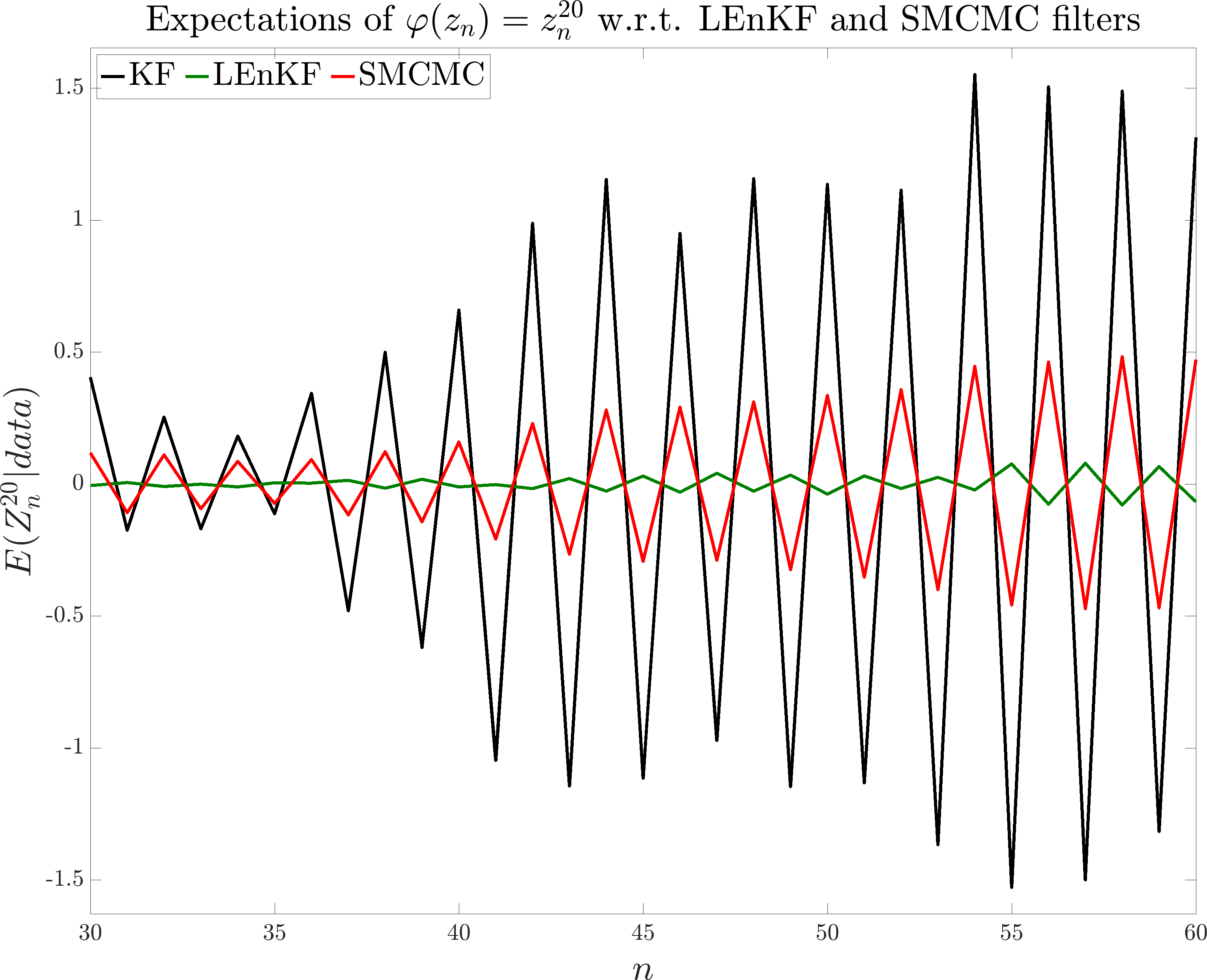}
\caption{ {(Model in Section \ref{sec:linear_model2}) This illustrates the means of the KF (truth), LEnKF and SMCMC for the \emph{observed} state variable $Z_n^{(20)}$ for all $n \in \{30,\cdots,60\}$.}}
\label{fig:smcmc_vs_lenkf_coord20}
\end{minipage}
\end{figure}

\begin{figure}[h!]
\centering
\includegraphics[scale=0.195]{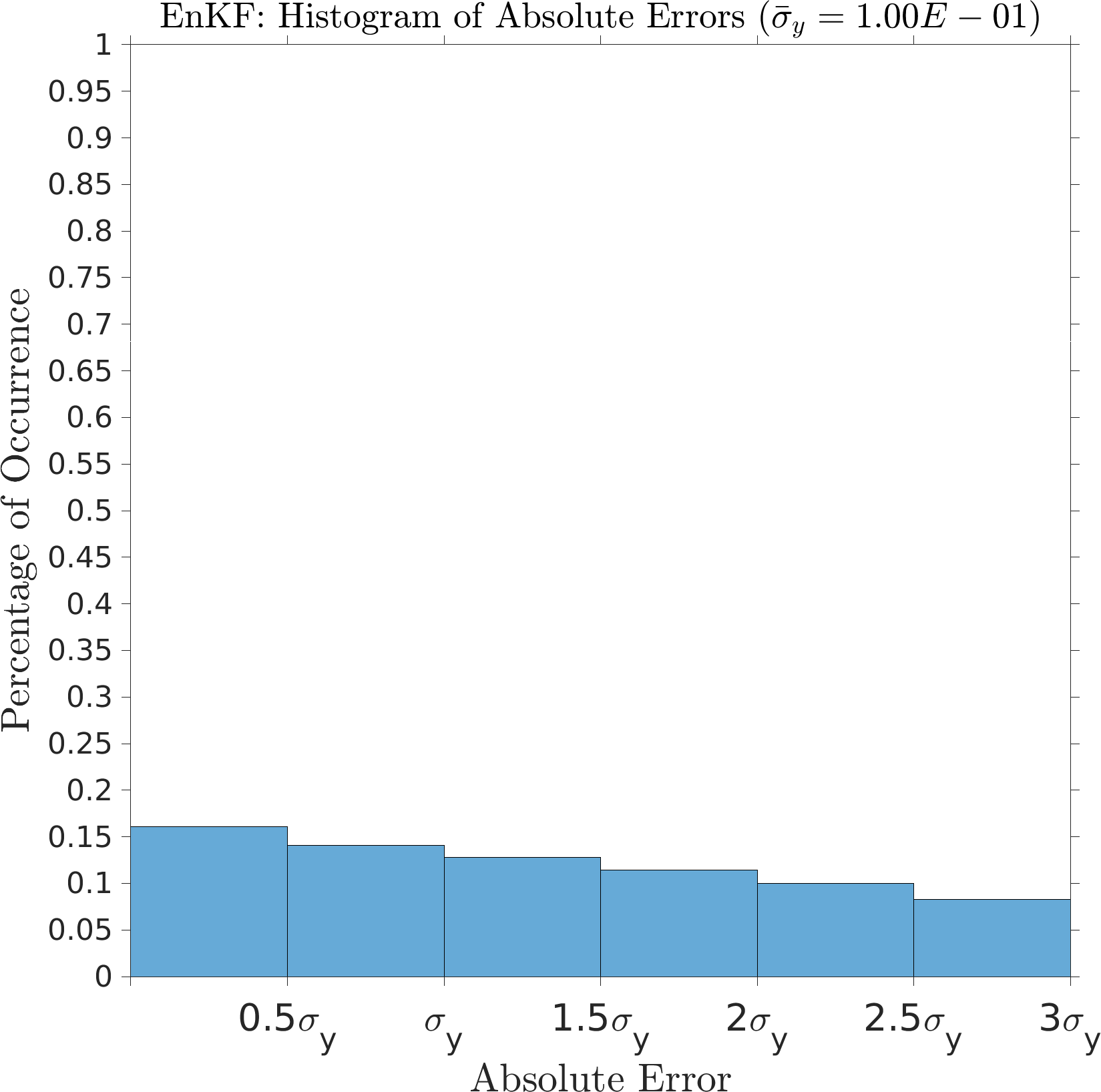}
\includegraphics[scale=0.195]{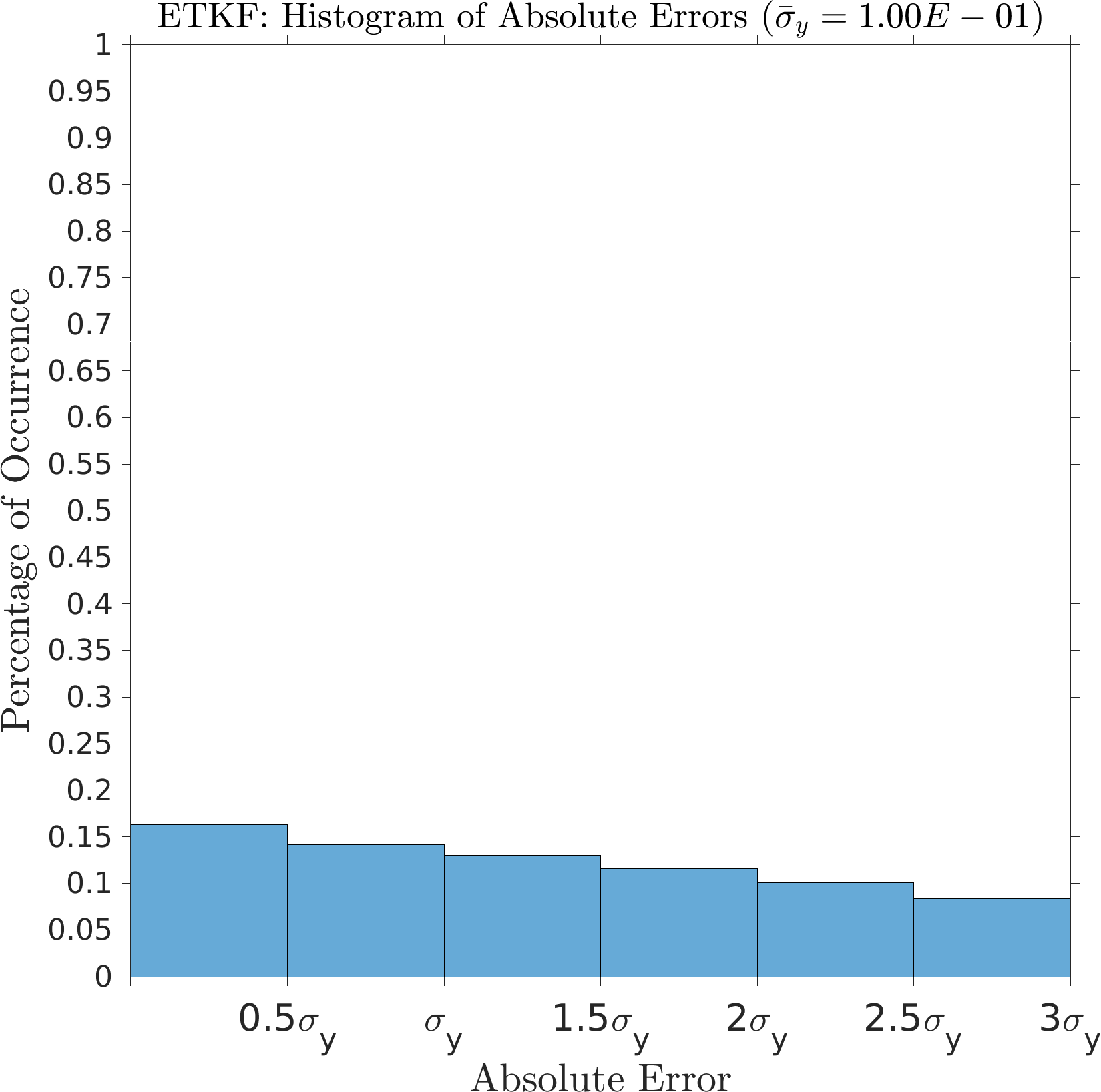}
\includegraphics[scale=0.195]{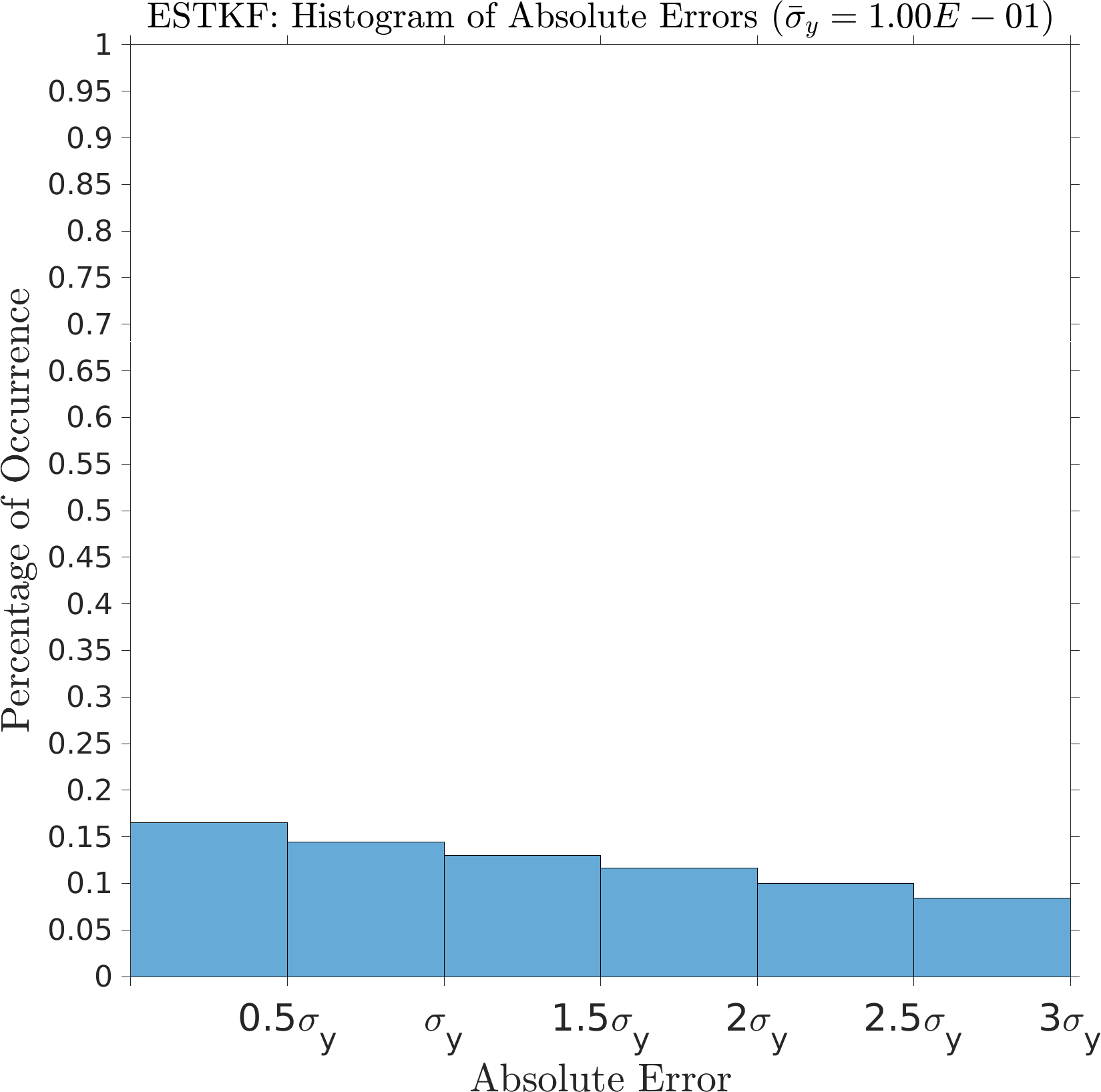}\\
\vspace{0.5cm}
\includegraphics[scale=0.195]{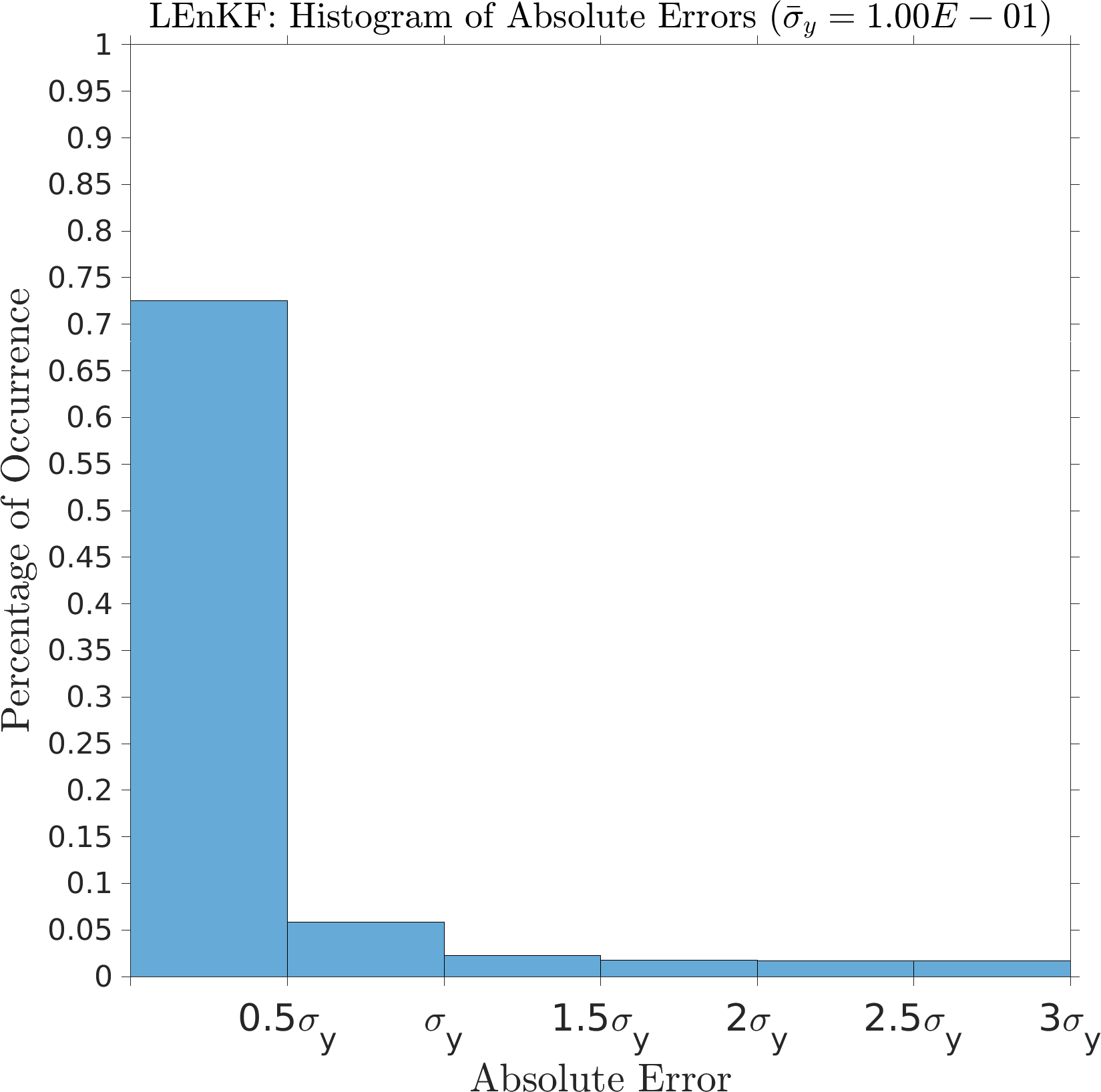}
\includegraphics[scale=0.19]{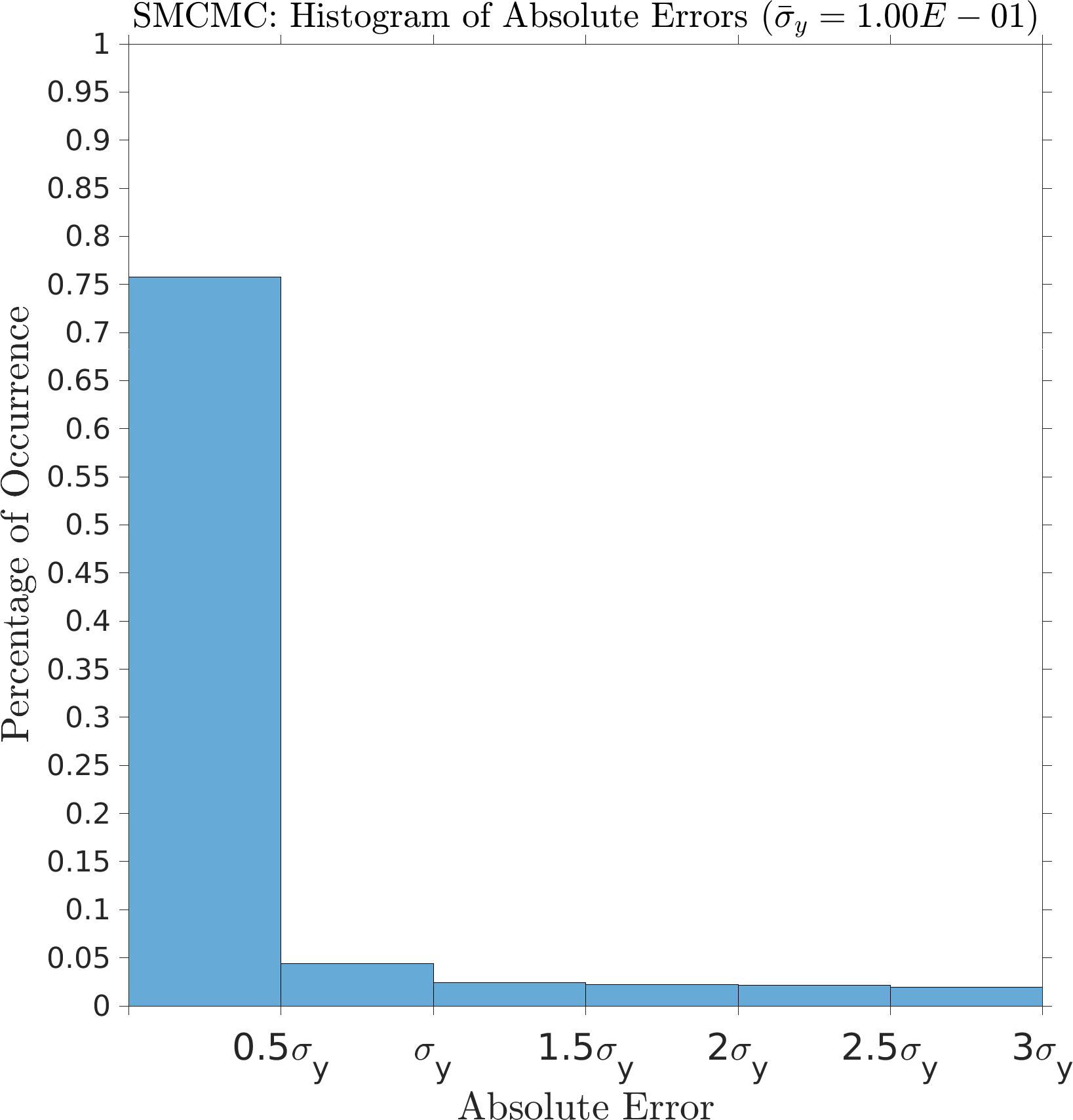}
\caption{ {(Model in Section \ref{sec:linear_model2}.) Histogram of absolute errors $|\text{Filter}_{\mu} - \text{KF}_{\mu}|$ at all state variables and at all times. Here we take $d=1600$ and $d_y=400$. The percentage of occurrence here is defined as the number of elements in the bin divided by the total number of elements $d \times (T+1)$.}}
\label{fig:Linear2_histo}
\end{figure}

 {
In Figures \ref{fig:smcmc_vs_lenkf_coord2_all}--\ref{fig:smcmc_vs_lenkf_coord20}, we plot the means of the filters when $d=1600$ and $d_y=400$ for observed and unobserved state coordinates. Figure \ref{fig:smcmc_vs_lenkf_coord2_all} shows the means of all filters for the unobserved state variable $Z_n^{(2)}$, where we can see that the non-localized ensemble filters are very inaccurate compared to the truth. In Figure \ref{fig:smcmc_vs_lenkf_coord2}, we omit the non-localized ensemble filters and keep only LEnKF and SMCMC to show that both methods are doing better than the rest when the state variable is unobserved. Figure \ref{fig:smcmc_vs_lenkf_coord4} shows that LEnKF gives a more accurate estimate for the observed state coordinate $Z_n^{(4)}$ compared to SMCMC filter, however, Figure \ref{fig:smcmc_vs_lenkf_coord20} shows the opposite when the observed coordinate is $Z_n^{(20)}$. This is also illustrated in the histograms in Figure \ref{fig:Linear2_histo}. As we can see from the histograms, the errors values are much higher for the non-localized ensemble methods because they fail to accurately estimate the filter for the unobserved state variables. }

\begin{figure}[h!]
\centering
\includegraphics[scale=0.4]{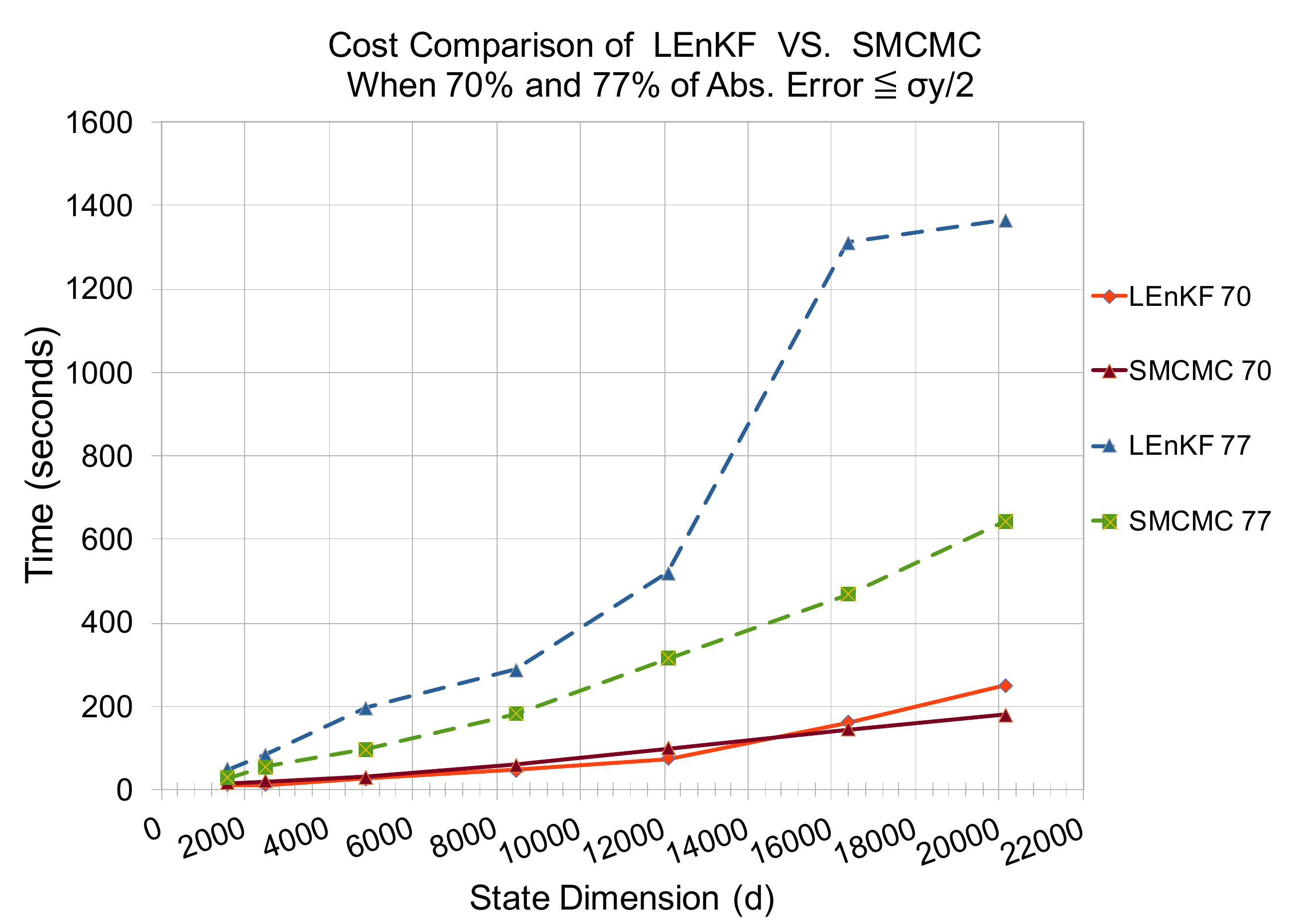}
\caption{ {(Model in Section \ref{sec:linear_model2}) We plot the cost (in seconds) for SMCMC and LEnKF methods for different state dimensions $d$. Two cases are considered: When 70\% of the absolute errors are less than $\sigma_y/2$ (solid lines), and when 77\% of the absolute errors are less than $\sigma_y/2$ (dashed lines).}}
\label{fig:smcmc_vs_lenkf}
\end{figure}

 {Next we compare between SMCMC and LEnKF in terms of accuracy and cost against the state dimension $d$. In Figure \ref{fig:smcmc_vs_lenkf}, we plot the cost, measured in seconds, of both methods in two cases: First, when 70\% of the absolute errors are less than $\sigma_y/2$. Second, when the requirement for accuracy is higher, that is, when 77\% of the absolute errors are less than $\sigma_y/2$. To get a higher accuracy we increase the \textcolor{black}{ensemble size} for LEnKF and the number of independent repeats $M$ for SMCMC (while fixing $N$ to $500$ and $N_{\text{burn}}$ to 100). In the first case, LEnKF runs slightly faster when $d = 1600, 2500, 4900, 8464, 12100$. However, when $d$ becomes larger, i.e. $d=16384 ~\& ~ 20164$, SMCMC runs faster. As for the second case, we had to use a large \textcolor{black}{ensemble size} for LEnKF to make the accuracy higher (for example we used $N=6000$ \textcolor{black}{ensemble members} when $d=20164$), which resulted in making LEnKF much more expensive to run than SMCMC (we run $M=350$ independent simulations, \textcolor{black}{each with $N=500$}, when $d=20164$).
}

\subsection{Rotating Shallow-Water Model Observed at Known  Locations}
\label{subsec:rSWE_known}
We consider a signal of the type in Example \ref{exam:pde}, where the PDE is associated to the shallow water equations (SWEs). The SWEs we consider are as follows
\begin{align*}
&\frac{\partial \zeta}{\partial t} + \frac{\partial (\eta u)}{\partial x} + \frac{\partial (\eta v)}{\partial y} = 0,\\
&\frac{\partial (\eta u)}{\partial t} + \frac{\partial}{\partial x} (\eta u^2 +\tfrac{1}{2}g \eta^2) + \frac{\partial (\eta uv)}{\partial y}  = g \eta\frac{\partial H}{\partial x} + f_1 \eta v,\\
&\frac{\partial (\eta v)}{\partial t} +  \frac{\partial (\eta uv)}{\partial x}  + \frac{\partial}{\partial y} (\eta v^2 +\tfrac{1}{2}g \eta^2) = g\eta\frac{\partial H}{\partial y} - f_1\eta u,
\end{align*} 
where $(x,y)\in[\underline{L}_x,\bar{L}_x]\times[\underline{L}_y,\bar{L}_y]$, $g$ is the gravitational acceleration, $f_1$ is the Coriolis parameter that is assumed to be varying linearly with $y$ such that $f_1=f_0 +\beta (y-y_0)$, 
where $f_0 = 2\Omega\sin \psi_0$, $\Omega=7.29\times 10^{-5} sec^{-1}$ the rate of earth's rotation, $\psi_0$ the central latitude of the region under study, $y_0$ is the $y$-value at $\psi_0$, and $\beta$ the meridional gradient of Coriolis force at $\psi_0$. Here $\eta$ represents the depth of the water (sea free surface to sea bottom), $H$ is the bathymetry which is defined as the distance from the geoid to the sea bottom (positive downwards), $\zeta$ is the elevation of the free surface measured from the geoid (positive upwards), therefore, $\eta=\zeta+H$, $u$ and $v$ are the horizontal velocities in $x$ and $y$ directions, respectively. The boundary conditions will be provided by the actual oceanographic data (based on a separate data assimilation procedure) and will be time varying. 

We use the finite-volume (FV) solution of the SWE {\cite{kurganov, zeitlin} which comprises of a 2-stage Runge-Kutta method combined with a local Lax-Friedrichs FV scheme}, with time step $\tau_k = (t_k -t_{k-1})/L$, hence, $t_k = kL$, with $t_0=0$ and its output will consist of $(Z_t)_{0\leq t \leq T}$, for some time $T>0$. The details are presented in Appendix \ref{sec:num_pde_solver}.
Let $N_x, N_y\in \mathbb{N}$ be the number of the cells in the grid in $x$ and $y$ direction respectively with $\Delta_x $, $\Delta_y$ the corresponding step sizes. The hidden signal at time $ t= t_{k-1} + l\tau_k\in [t_{k-1}, t_k)$, is the vector given by
\begin{equation*}
Z_t = [(\eta_{i}^t)_{1\leq i \leq N_xN_y }, (u_{i}^t)_{1\leq i \leq N_xN_y },(v_{i}^t)_{1\leq i \leq N_xN_y } ]^\top \in \mathbb{R}^{3N_xN_y},
\end{equation*}
where $Z_0$ is known and $(\eta_{i}^t)_{1\leq i \leq N_xN_y }, (u_{i}^t)_{1\leq i \leq N_xN_y },(v_{i}^t)_{1\leq i \leq N_xN_y }$ are row vectors obtained from the approximate solver detailed in Appendix \ref{sec:num_pde_solver},
$\mathsf{Z}:=\mathbb{R}^{3N_x N_y}$.

At prescribed times $(t_k)_{k\geq 1}$ we add Gaussian noise, $(W_{t_k})_{k\geq 1}$, to the output of the numerical solution of the PDE. To preserve the boundary conditions the noise is constructed such that it is zero at the boundary. In  particular,  for $\eta^{t_k}$ and $k\in \mathbb{N}$, we use 
\begin{equation}
    [\Xi^{\eta}_{t_k}]_{l=1,s=1}^{N_y,N_x}= \sum_{i=0}^{J-1} \sum_{j=0}^{J-1}\epsilon_{t_k}^{\eta, (i,j)} \sin\left(\frac{\pi j y_l}{\bar{L}_y-\underline{L}_y}\right) \sin\left(\frac{\pi i x_s}{\bar{L}_x-\underline{L}_x}\right),
\end{equation}
and similarly for $u^{t_k},v^{t_k}$ design $\Xi_{^{t_k}}^u,\Xi_{^{t_k}}^v$
and then vectorize to get
\begin{equation*}
W_{t_k}=[\text{Vec}(\Xi_{^{t_k}}^\eta)^T, \text{Vec}(\Xi_{^{t_k}}^u)^T,\text{Vec}(\Xi_{^{t_k}}^v)^T]^T.
\end{equation*}
Here $J\in\mathbb{N}$ is a user chosen number of Fourier modes, $\epsilon_{t_k}^{\cdot,(i,j)} \sim \mathcal{N}(0,\sigma^2/(i\vee j+1))$, for $i,j \in \{0,\cdots,J-1\}$, where $i\vee j$ here means the maximum of $\{i,j\}$, and $\sigma > 0$, see Appendix \ref{appx:noise} for the specific implementation approach. \textcolor{black}{In this example we have that $d=3N_xN_y$ and matrix-vector calculations have been carried out at a cost of $\mathcal{O}(d^2)$.  We note that the overall algorithm could have been implemented using FFT at a cost $\mathcal{O}(N L d \log d)$ per observation time with $L$  being the total time discretization steps used in each $k$ as per Example \ref{exam:xbar}. }

The observations in this model are obtained from a set of $N_d$ drifters in the region of study that are assumed to be moving according to the velocity components of the {solution of the} SWE above. {The observations (and their locations) are generated from the signal before running the filtering algorithm}. The observation model is taken as
\begin{equation*}
Y_{t_k} = \mathscr{O}_{t_k}(Z_{t_k}) + V_{t_k}, \quad V_{t_k} \stackrel{\textrm{i.i.d.}}{\sim} \mathcal{N}_{d_y}(0,\sigma_y^2 I_{d_y}),\quad t_k \in \mathsf{T},
\end{equation*}
where $\mathscr{O}_{t_k}:\mathsf{Z}\to \mathsf{Y}^{N_d}$ is an $\mathbb{R}^{d_y}$-vector valued function containing measurements of $(u_{i}^{t_k})$ and $(v_{i}^{t_k})$ from the signal $Z_{t_k}$ at time $t_k$ (we do not measure $\eta^{t_k}$). These are collected from drifters whose positions move according to a kinematic model, where in \eqref{eq:x_tilde} we use the values from the velocity fields at each location and therefore set $h(x^j_{t_k},Z_{t_k})=[u^{t_k}(x_{t_k}^j),v^{t_k}(x_{t_k}^j)]$, $j \in \{1,\cdots,N_d\}$.
Due to the space discretization involved in the PDE, in practice the observation location is chosen as the closest point on the grid at each time. In Figure \ref{fig:obs_loc}, we present the basic idea of locating the closest point on the grid via an illustration. 
First we locate the four grid points surrounding each drifter based on its location (according to \eqref{eq:x_tilde} in red).
Then we pick the closest grid point to each drifter. The set of picked grid nodes will correspond to the approximate spatial locations of the observations {(see Figure \ref{fig:obs_loc})} and similarly for the set of observations are the values of $(u_{i}^{t_k})$ and $(v_{i}^{t_k})$. 
Compared to the ideal physical quantities this observation model contains a discretization error that approaches zero as $\Delta_x, \Delta_y \to 0$.

\subsubsection{Simulation Settings}
The region of simulation is a domain of the Atlantic Ocean restricted to the longitude and latitude intervals $[-51^\circ, -41^\circ]$, $[17^\circ, 27^\circ]$, respectively. We use Copernicus Marine Services \cite{coper} to obtain the bathemetry $H$, the sea surface height above geoid $\overline{\eta}$, and the horizontal velocities $\overline{u}$ \& $\overline{v}$, with a $1/12$ degree horizontal resolution, at an hourly rate from 2020-03-01 to 2020-03-05. The values of $\overline{\eta}$, $\overline{u}$ \& $\overline{v}$ at time 00:00 2020-03-01 is used as an initial state $z_0$. As for the boundary conditions, we also use the values of $\overline{\eta}$, $\overline{u}$ \& $\overline{v}$ (at the boundaries) interpolated in time. For the observations, we assume that there are $N_d =12$ drifters in the region during the period of simulation that observe $u$ and $v$. We obtain the drifters initial locations from the set of data available in \cite{adam1,adam2}. Drifters stay at the surface of the water and move with surface currents and their positions are tracked every 10 minutes to 1 hour (depending on the programme) and they can move up to $2m/s$, so up to $100km$ a day on e.g. the Gulf Stream, but typically indeed it is much less ($<10km$) especially away from the boundary currents. They provide a Lagrangian horizontal velocity data near the surface (thus $d_y= 2 N_d = 24$) at roughly hourly resolution. Then, 26 independent simulations of Algorithm \ref{alg:1} were run in parallel with the following parameters: $N_x = 121$, $N_y=121$, $d = 3N_xN_y = 4.3923 \times 10^4$, $\Delta_x =8.602 \times 10^3$ meters, $\Delta_y=9.258\times 10^3$ meters, $\tau_k = 60$  seconds for all $k\in \mathbb{N}$, $T=7.2 \times 10^6$ seconds (i.e. for 33.3 hours), $L=10$, $J=8$, $N=1200$, $N_{burn}=200$, $\sigma = 2\times 10^{-4}$, $\sigma_y=1.45 \time 10^{-2}$.

\subsubsection{Results with synthetic drifter locations}
\begin{figure}[h!]
\begin{minipage}[b]{.46\textwidth}
\centering
\includegraphics[trim={7cm 17cm 7cm 5.5cm},clip, scale=0.5]{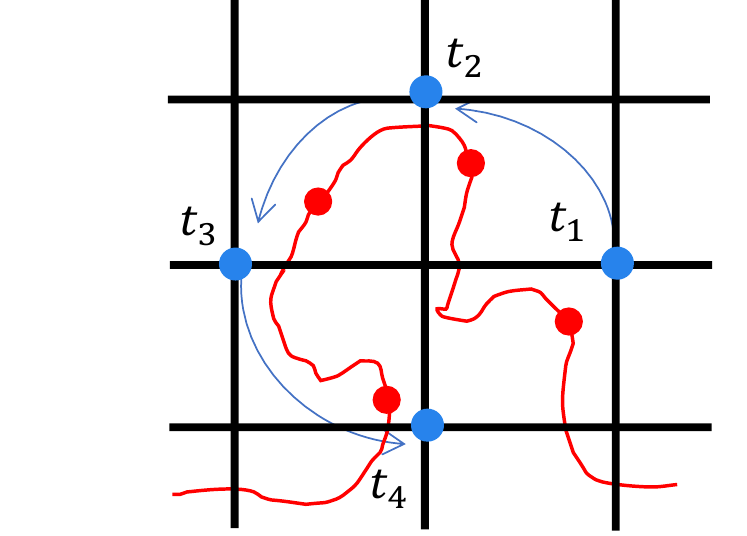}
\caption{{The illustration depicts the process of selecting which state variables to be observed based on the drifter's location. The red circles represent a drifter at various points in time, and the red curve indicates its track. The blue circles correspond to the nearest surrounding grid point at the times of observation.}}
\label{fig:obs_loc}
\end{minipage}
\hfill
\begin{minipage}[b]{.46\textwidth}
\centering
\includegraphics[scale=0.15]{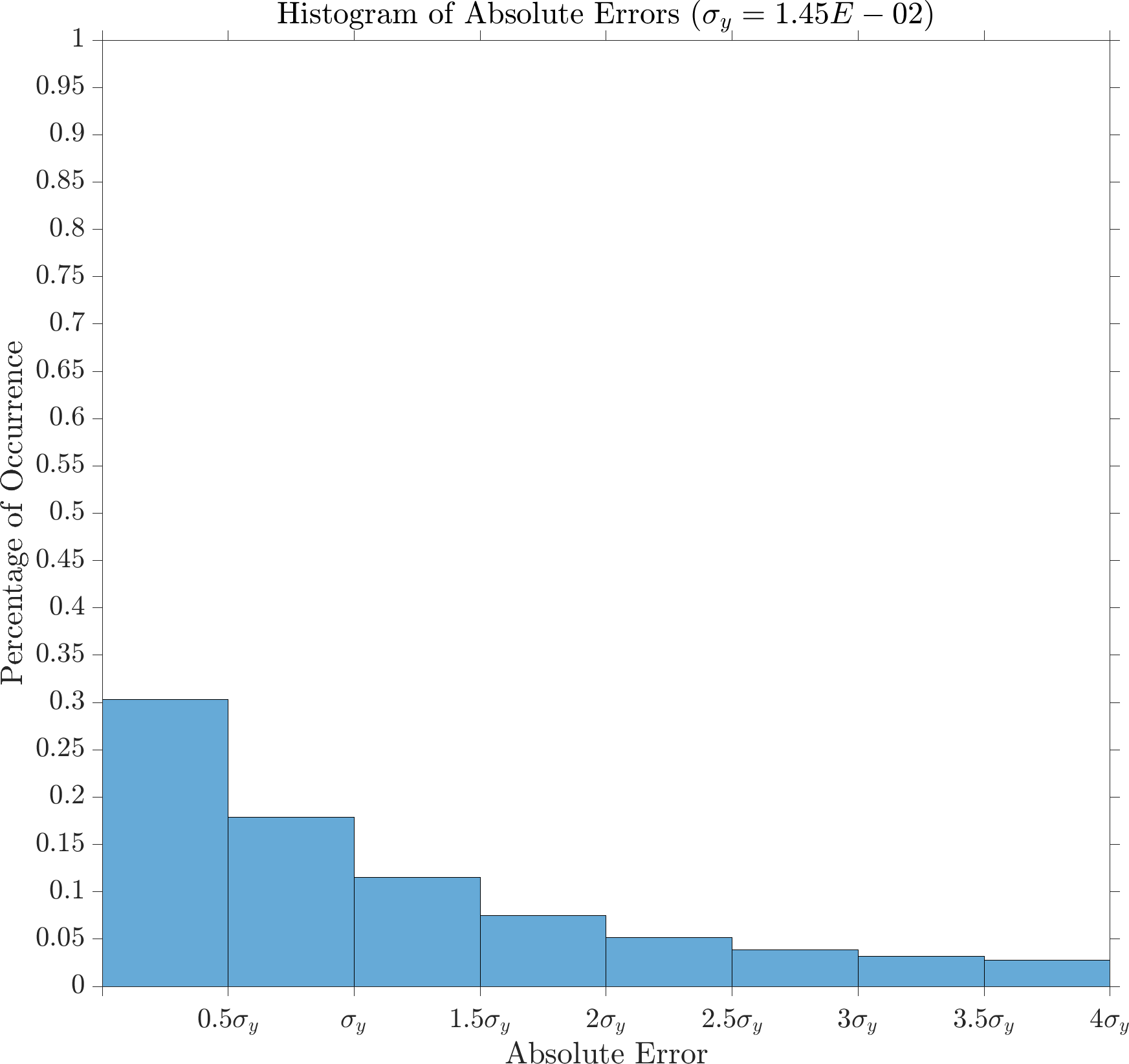}
\caption{(Known locations example) Histogram of absolute differences: $|\text{Filter Mean} - \text{Signal}|$ at all state variables and at all times.  }
\label{fig:hist_errors_known}
\end{minipage}
\end{figure}

\begin{figure}[h!]
\centering
\hspace*{-2cm} 
\includegraphics[scale=0.27]{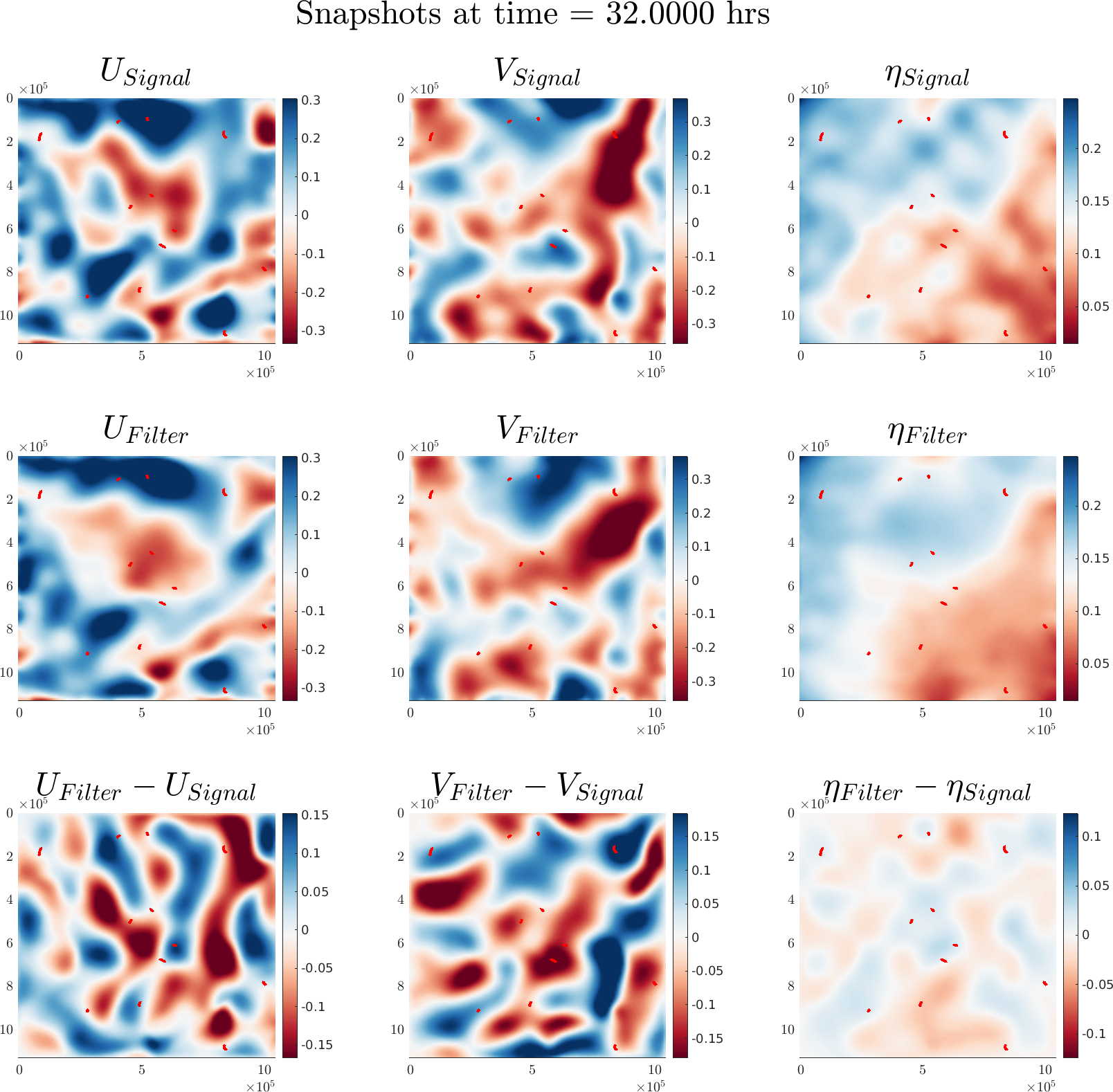}
\caption{Snapshot of simulation at time $=$ 32hrs for the SW model with observations of known spatial locations. We present the hidden signal, the filter mean and their difference for the rotating shallow-water HMM with observations of known locations. Red curves illustrate the tracks of drifters in the region, which are moving according to the signal.}
\label{fig:known_loc_last}
\end{figure}

In Figure \ref{fig:hist_errors_known} we show a histogram of the absolute errors, defined as the absolute difference between the values of the filter and the \emph{hidden signal} ({from which the data is generated}) at all state variables and all times. Furthermore, in Figure %\ref{fig:known_loc_first}-
\ref{fig:known_loc_last} we present a snapshots of the hidden signal, the filter and their difference after
%s 8hr, 20hr, 26hr and
 32hrs. Also shown in the snapshots the tracks of the drifters which were computed before implementing the filtering algorithm.  The ratio of the number of observations to the number of state variables is $24/43923 = 0.055\%$. Even with a small number of observations, the results show that the filtering technique presented here is quite effective. Finally, we note that these simulations took about \textcolor{black}{67 hours} to run 26 independent repeats on 52 cores.

%\begin{figure}
%\centering
%\hspace*{-2cm} 
%\includegraphics[scale=0.15]{figures/known_loc8.png}
%\caption{Snapshot of simulation at time $=$ 8hrs for the SW model with observations of known spatial locations. We present the hidden signal, the filter mean and their difference for the rotating shallow-water HMM with observations of known locations. Red curves illustrate the tracks of drifters in the region, which are moving according to the signal.}
%\label{fig:known_loc_first}
%\end{figure}

\begin{figure}[h!]
\centering
\hspace*{-2cm} 
\includegraphics[scale=0.27]{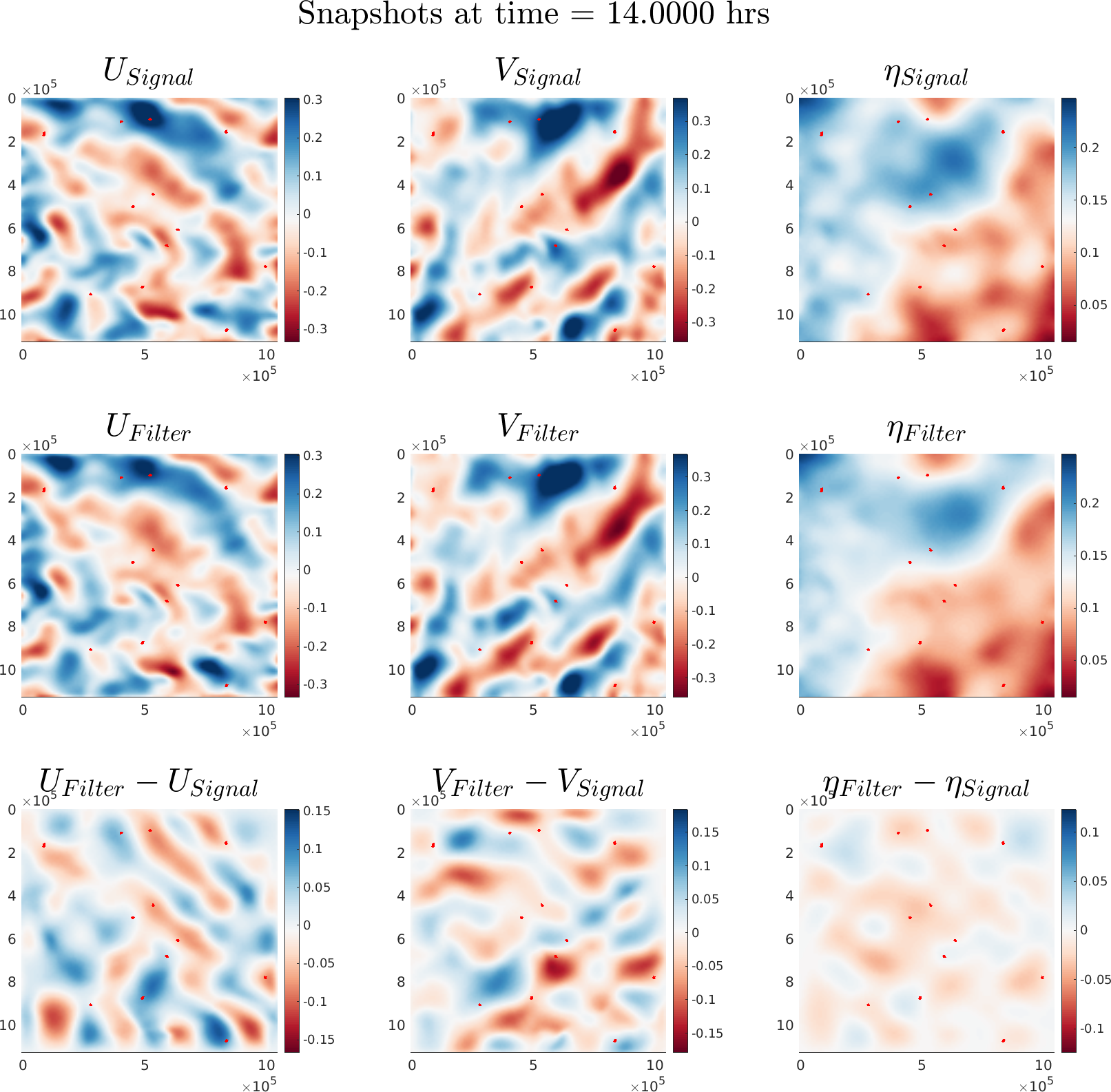}
\caption{Snapshot of simulation at time $=$ 14hrs for the SW model with observations of known spatial locations}
\end{figure}

\begin{figure}[h!]
\centering
\hspace*{-2cm} 
\includegraphics[scale=0.27]{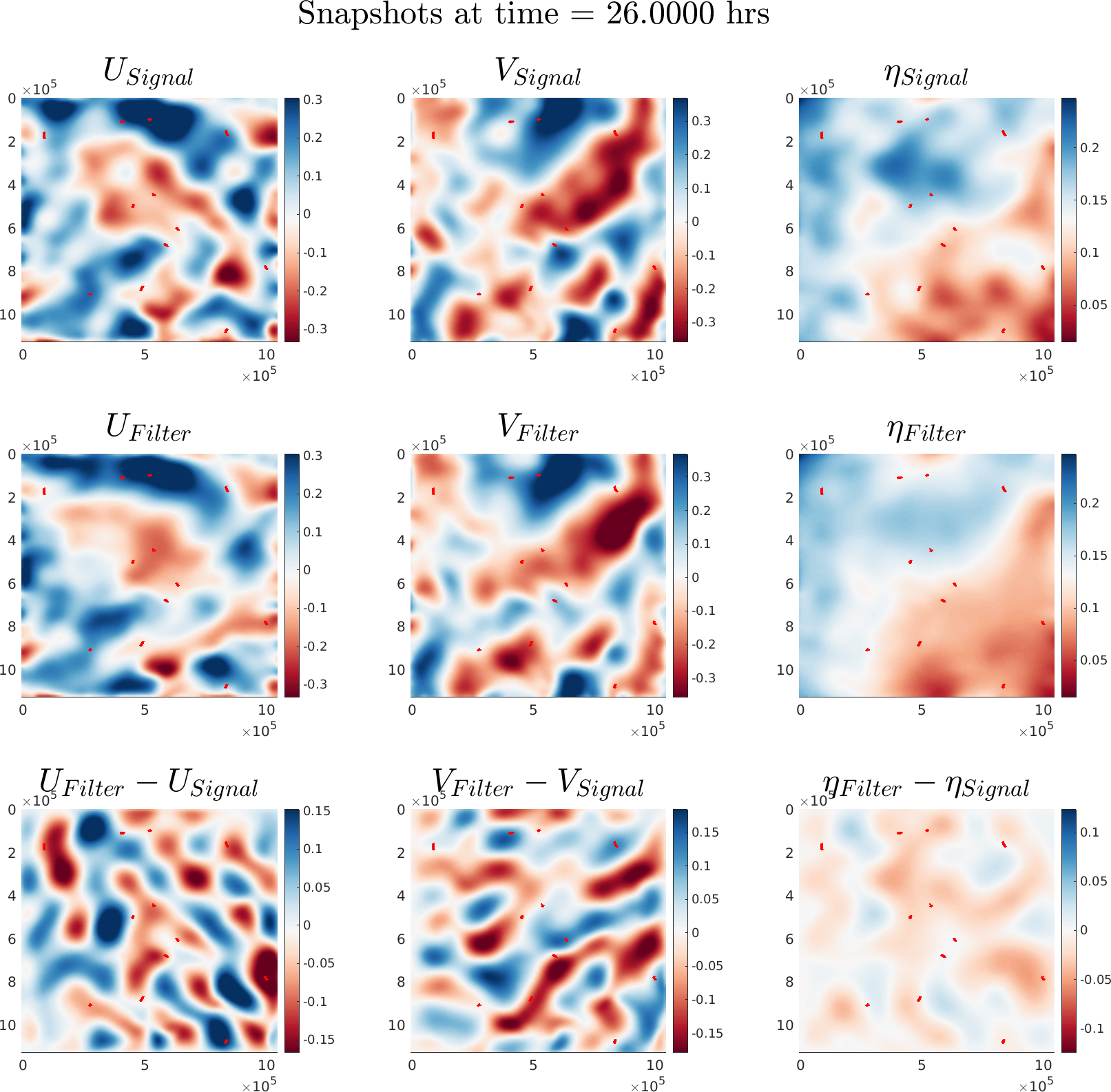}
\caption{Snapshot of simulation at time $=$ 26hrs for the SW model with observations of known spatial locations.}
\end{figure}

\subsection{Rotating Shallow-Water Model Observed at Unknown  Locations}

Here we consider the same model as in the previous section except that it is assumed that the spatial locations of the observational data are unknown and we use real drifter data from NOAA. For $t_k \in \mathsf{T}$, the set of observations is the measurements of $u$ and $v$ obtained by the drifters in the region of simulation at time $t_k$. To evaluate the function $G\left((z_{t_k},{\overline{x}}_{t_k}),{y}_{t_k}\right)$, which is a Gaussian density with mean $\mathscr{O}_{t_k}(z_{t_k})$ and a covariance matrix $\sigma_y^2I_{d_y}$, at the given observations at time $t_k$, we need to determine the mapping $\mathscr{O}_{t_k}:\mathsf{Z}\to \mathsf{Y}^{N_d}$. This is done the same way as in the previous example except now we use the estimate of the mean spatial locations of the drifters as in Example \ref{exam:xbar}.

\subsubsection{Simulation Setting}
The simulation region is the same as it was in the preceding example. \cite{adam1,adam2} provided the data used in this analysis, which showed the presence of 12 drifters in that region during the simulation period along with hourly measurements of $u$ and $v$ that we also extrapolated over time. The mean error of the measurements during the simulation period is computed over all times and all drifters and is denoted by $\overline{\sigma}_y=0.0145$, whereas the minimum error value is 0.0012 and the maximum error value is 0.0375. Here, we should point out two ways in which the data in this example differs from that in the preceding example: i) The data in this case is considered to have been obtained at unknown locations, whereas the data in the previous example was taken at known locations. ii) Second and most importantly, the data in this example is real measurements, whereas the one in the other example is synthetic. Then, using the same parameters as before we run 26 independent simulations of Algorithm \ref{alg:unknown} in parallel.

\subsubsection{Results with real data}

\begin{figure}[h!]
\centering
\includegraphics[scale=0.15]{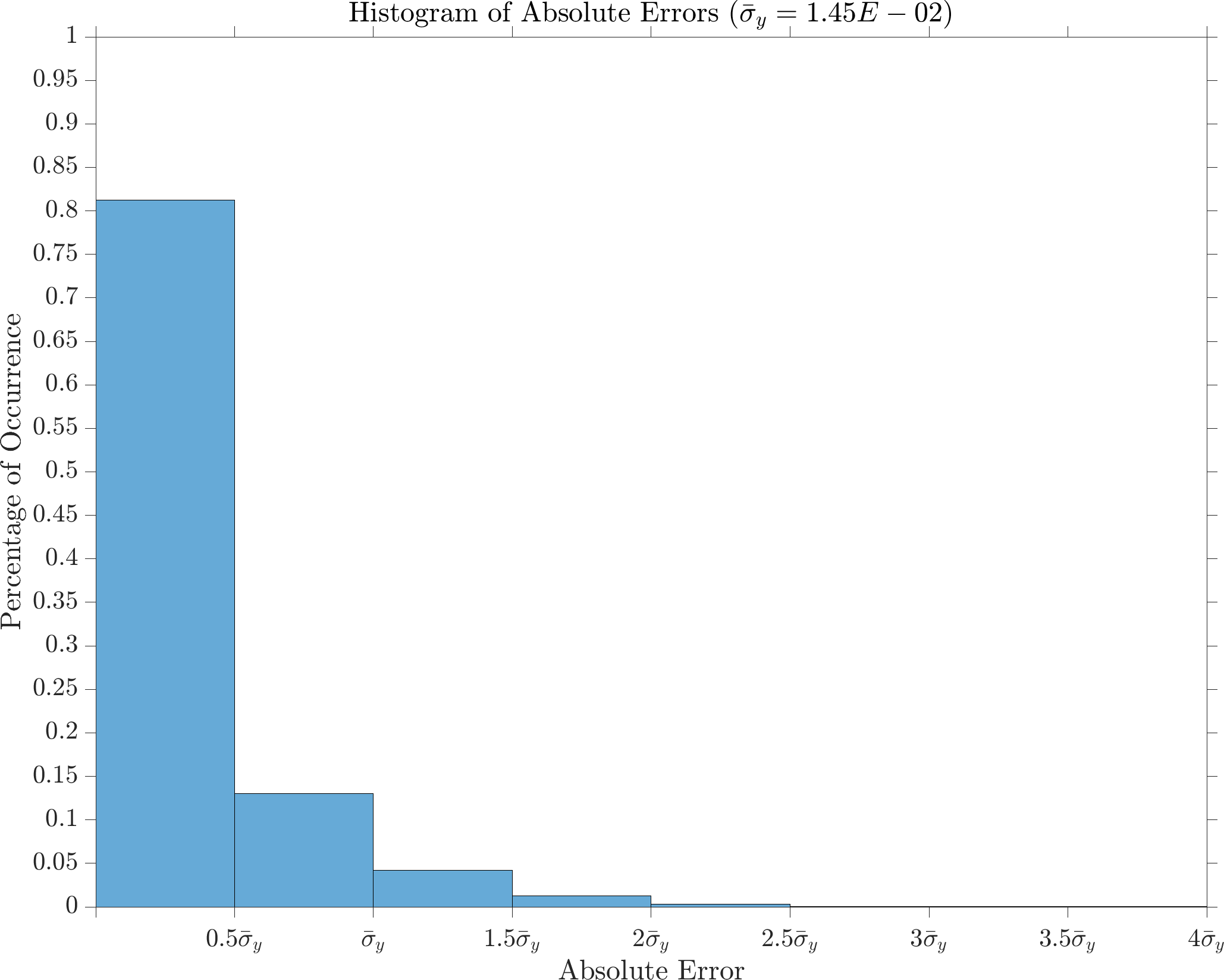}
\caption{(Unknown locations example) Histogram of learning differences: $|\text{Filter Mean} - \text{Prior}|$ at all state variables and at all times. The percentage of occurrence here is defined as the number of elements in the bin divided by the total number of elements $d \times (T+1)$. }
\label{fig:hist_errors_unknown}
\end{figure}

\begin{figure}[h!]
\centering
\hspace*{-2cm} 
\includegraphics[scale=0.27]{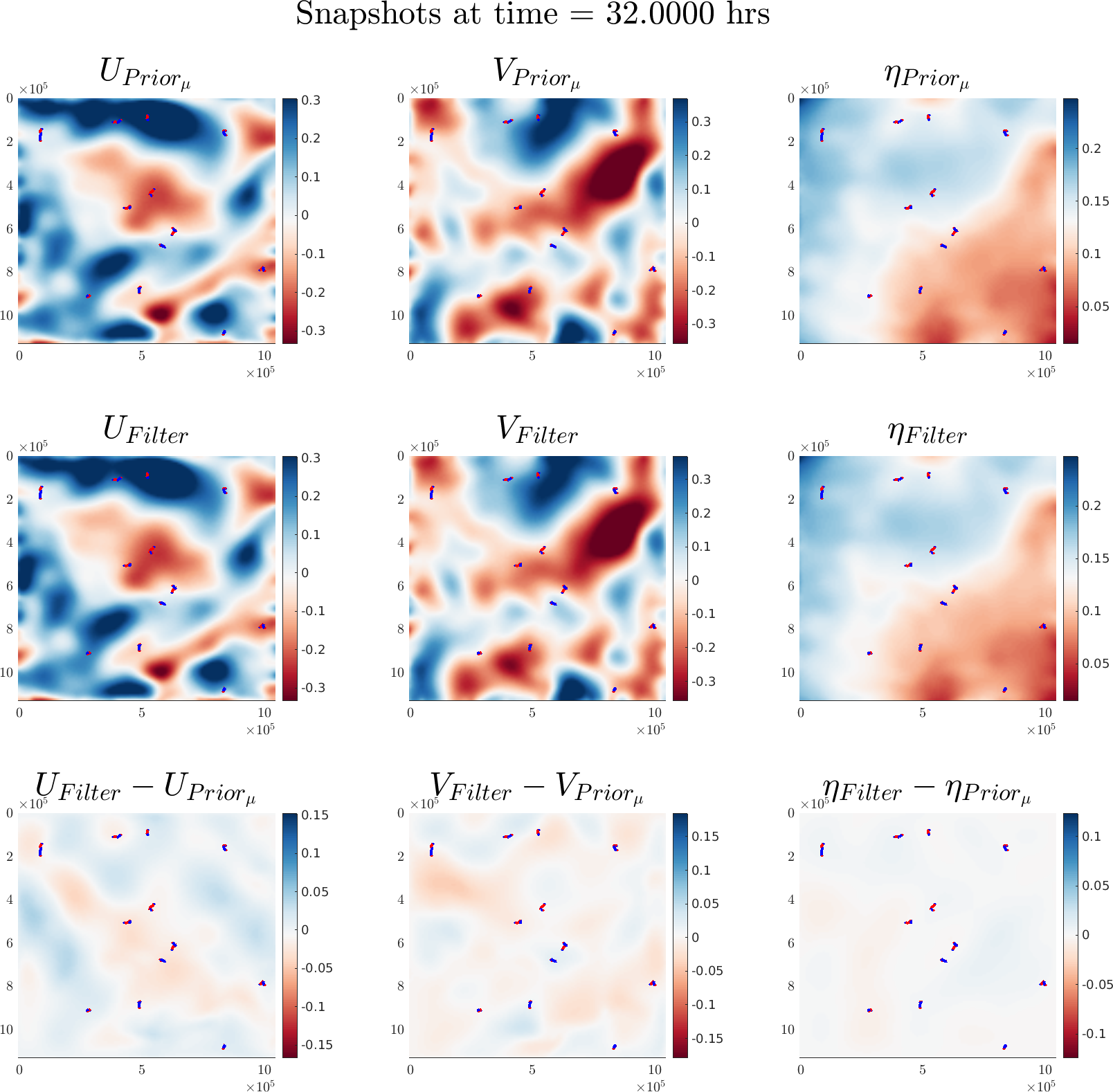}
\caption{Snapshot of simulation at time $=$ 32hrs for the SW model with observations of unknown spatial locations. We show the prior mean (reference signal), the filter mean and their difference for the rotating shallow-water HMM with observations of unknown spatial locations. The blue curves illustrate the tracks of the drifters according to the data from \cite{adam1}, while the red curves illustrate the mean of 50 tracks of the drifters that are moving according to the prior. The red and blue tracks are presented here only for the purpose of illustration; Algorithm \ref{alg:unknown} does not use the blue tracks but aims to estimate them. See also Figure \ref{fig:zoomed_drifters} in which the region $[4\times 10^{5}, 7\times 10^{5}] \times [4\times 10^{5}, 7\times 10^{5}]$ is zoomed-in to show the difference between the blue and red drifters' tracks.}
\label{fig:unknown_loc_last}
\end{figure}

In Figure \ref{fig:hist_errors_unknown} we show a histogram of the absolute errors, defined as \textcolor{black}{the absolute difference between the filter mean and the reference signal}. The histogram shows that 93.6\% of the filter values are within $
\overline{\sigma}_y/2$ from the reference signal values. All comparisons are with a {\bf reference} signal that approximates the mean of the prior distribution. In this example is taken as the mean of 50 independent runs of the SW dynamics with noise using the same initial value $Z_0$ and the same boundary conditions. Furthermore, in Figure \ref{fig:unknown_loc_last} %-\ref{fig:unknown_loc_last} 
we present a snapshots of the hidden signal, the filter and their difference after %s 8hr, 20hr, 26hr and 
32hrs. Also shown in the snapshots the tracks of the drifters in red and blue. Red tracks refer to the mean spatial locations of the drifters computed according to the reference signal, whereas the blue tracks refer to the spatial locations obtained from the data set in \cite{adam1}. The latter are not used by the algorithm, which manages to estimate them on the fly. 

%\begin{figure}
%\centering
%\hspace*{-2cm} 
%\includegraphics[scale=0.15]{figures/unknown_loc8.png}
%\caption{Snapshot of simulation at time $=$ 8hrs for the SW model with observations of unknown spatial locations. We show the prior mean (reference signal), the filter mean and their difference for the rotating shallow-water HMM with observations of unknown spatial locations. The blue curves illustrate the tracks of the drifters according to the data from \cite{adam1}, while the red curves illustrate the mean of 50 tracks of the drifters that are moving according to the prior. The red and blue tracks are presented here only for the purpose of illustration; Algorithm \ref{alg:unknown} does not use the blue tracks but aims to estimate them.}
%\label{fig:unknown_loc_first}
%\end{figure}

A comparison of the \textcolor{black}{absolute differences} (third row) in Figure \ref{fig:known_loc_last}
and in the known observer trajectory case 
Figure \ref{fig:unknown_loc_last} cannot be made directly due to the differences in the reference and the hidden signal. In Figure 
%\ref{fig:unknown_loc_first}-
\ref{fig:unknown_loc_last}, it is notable that despite the lack of observer position trajectories, the difference between posterior and prior mean suggest informative likelihoods for $u,v$ and that the posterior does manage to gain significant information for these variables. In contrast for $\eta$ these differences are smaller and learning occurs via the dependence of $\eta$ with $u,v$ in the dynamics. The results for this example again show that the filtering technique presented here is quite effective even when the spatial locations of the observations are assumed unknown. Finally, we note that these simulations took around \textcolor{black}{68.5 hours} hours to run 26 independent repeats on 52 cores.

\begin{figure}[h!]
\centering
\hspace*{-2cm} 
\includegraphics[scale=0.27]{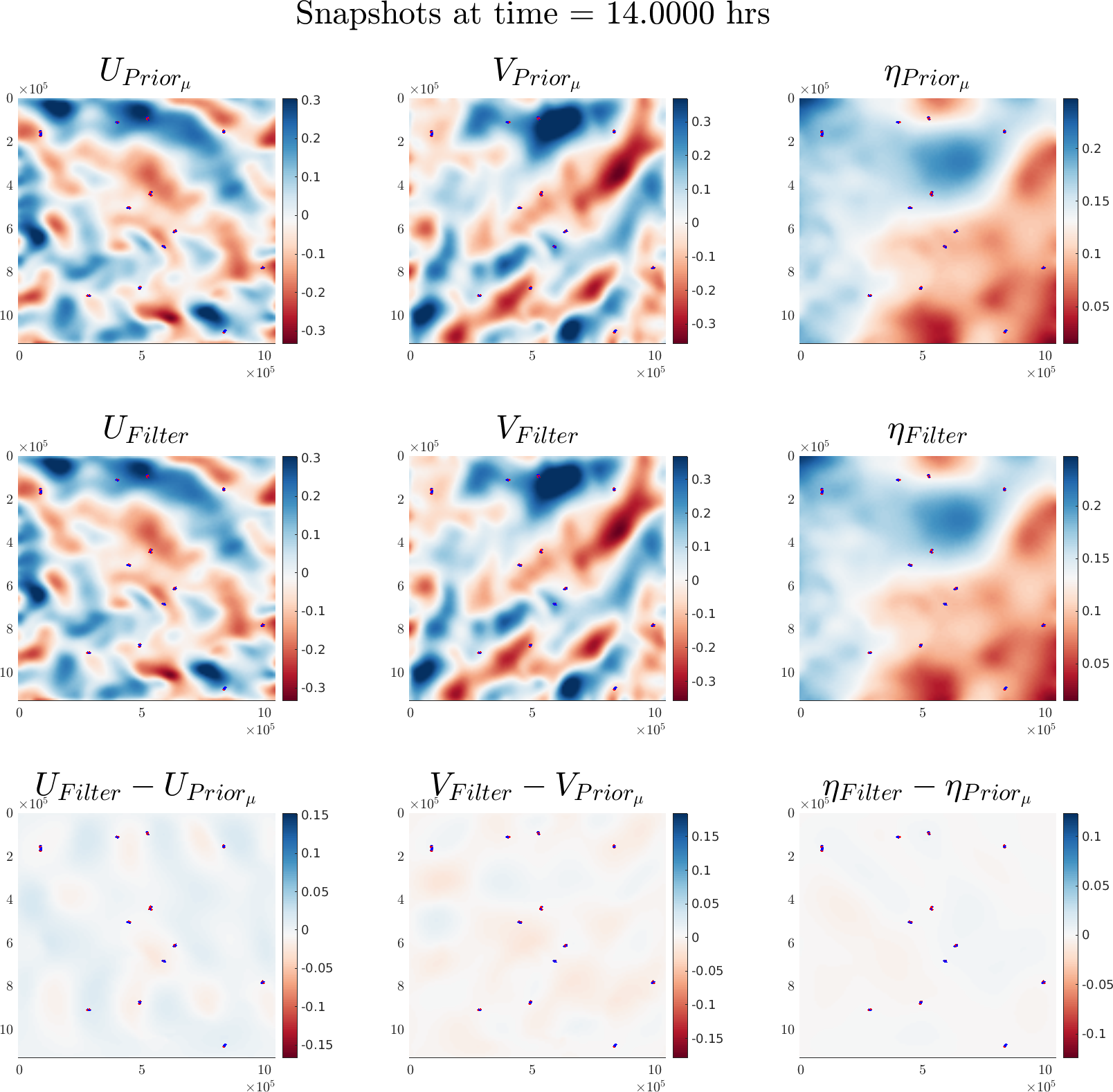}
\caption{Snapshot of simulation at time $=$ 14hrs for the SW model with observations of unknown spatial locations}
\end{figure}

\begin{figure}[h!]
\centering
\hspace*{-2cm} 
\includegraphics[scale=0.27]{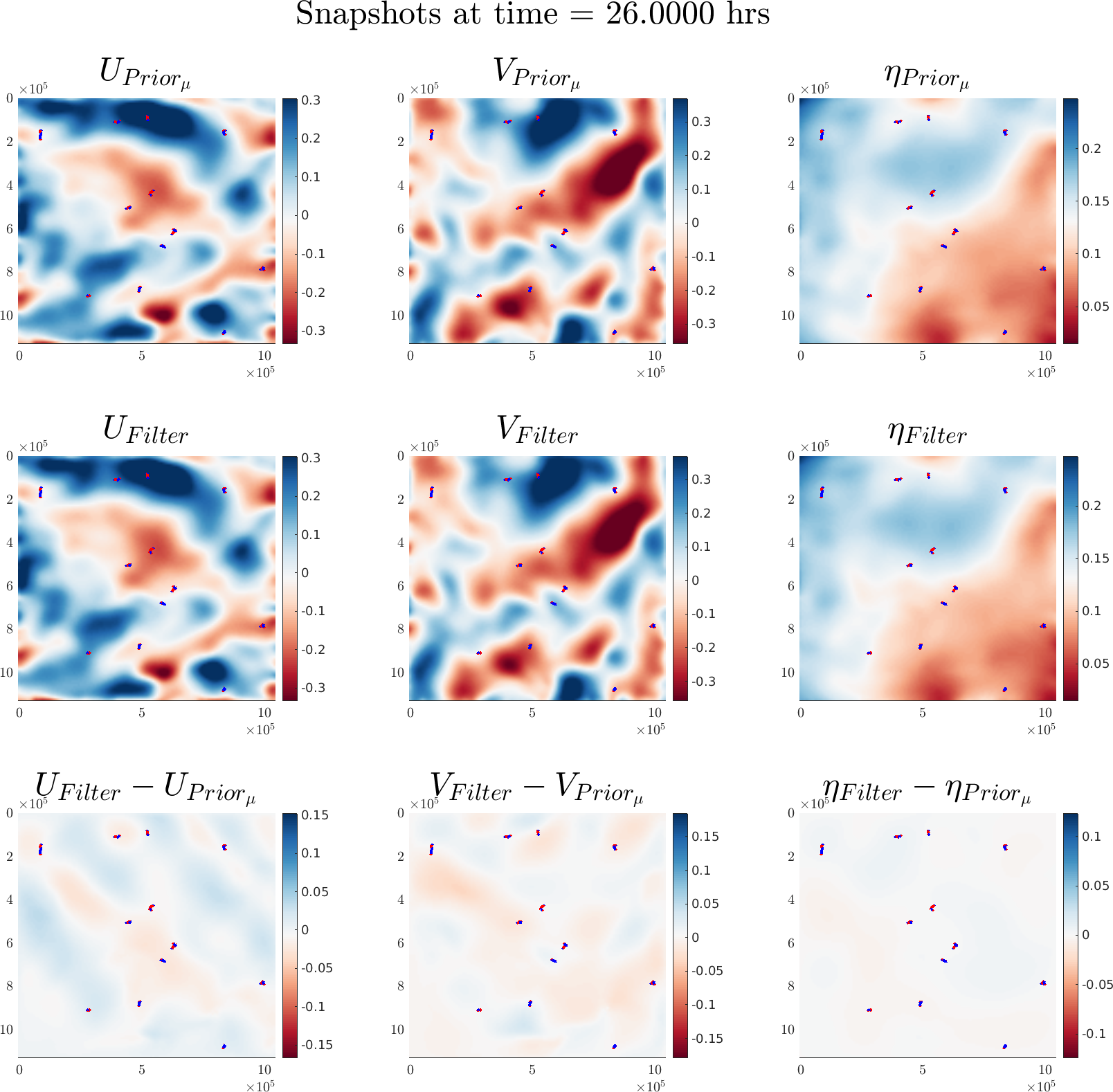}
\caption{Snapshot of simulation at time $=$ 26hrs for the SW model with observations of unknown spatial locations.}
\end{figure}

\begin{figure}[h!]
\centering
\hspace*{-2cm} 
\includegraphics[scale=0.15]{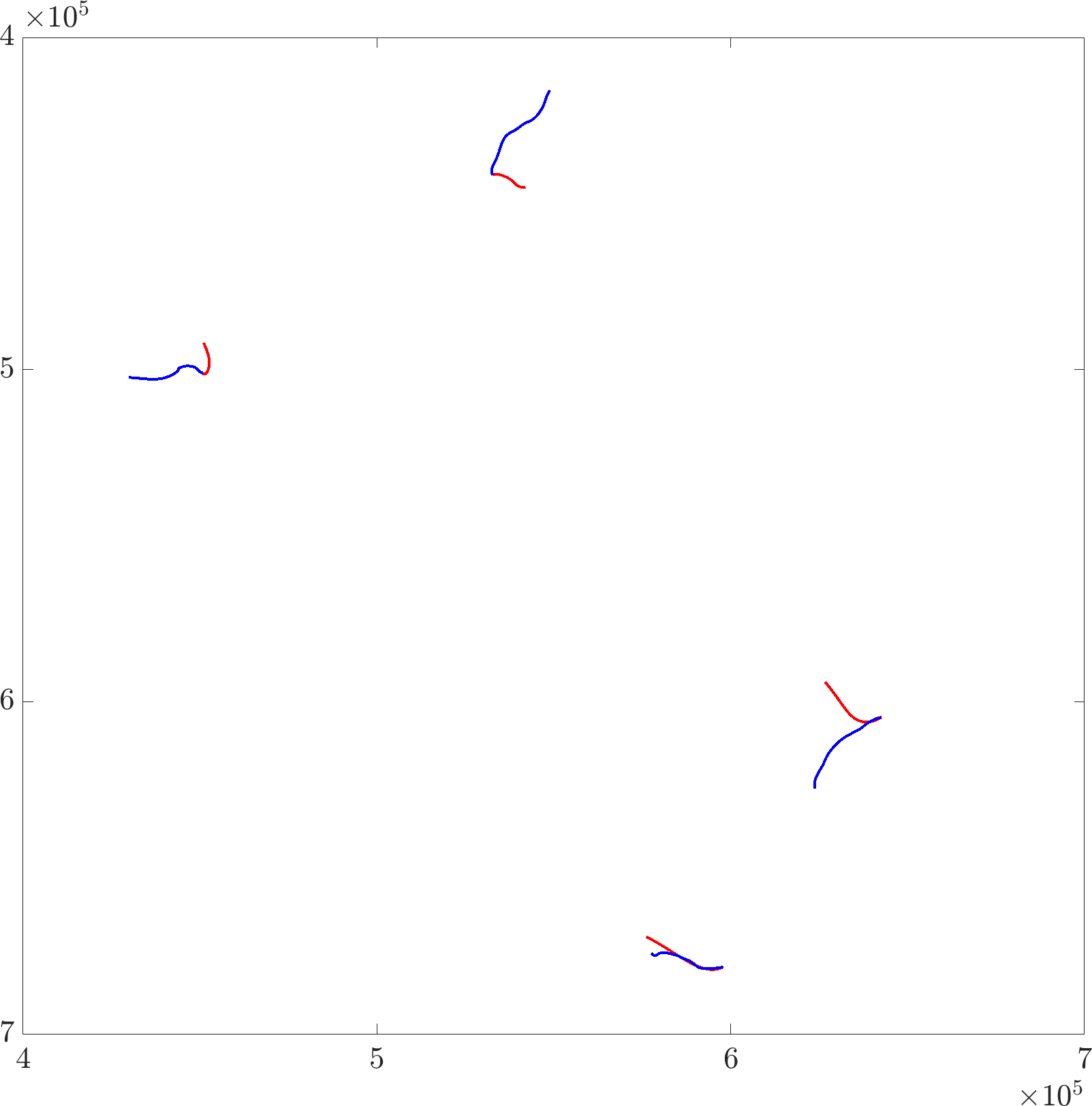}
\caption{This is the region $[4\times 10^{5}, 7\times 10^{5}] \times [4\times 10^{5}, 7\times 10^{5}]$ from Figure \ref{fig:unknown_loc_last} which contains four drifters is zoomed-in in order to show how different can be the blue (from real data) and red (from prior) tracks.}
\label{fig:zoomed_drifters}
\end{figure}

\section{Discussion}\label{sec:disc}

\textcolor{black}{In this article we have considered the adaptation of the SMCMC method in the context of a class of high-dimensional filtering problems. We showed in several examples that there is great promise in the application of this approach relative to some of the existing state-of-the-art numerical methods. In addition, the method is provably convergent in several contexts, even for non-linear problems. It should also be noted that the methodology that is presented is not restricted to the class of problems considered in this article and in general can be applied to many filtering problems as is evidenced in e.g.~\cite{cent}.  It is simple to extend the models to include static (in-time) and unknown parameters in the model, which are to be estimated using Bayesian methods. This estimation can be performed using the methods considered in this article.}

\textcolor{black}{As, naturally, SMCMC depends upon an MCMC kernel, the ability of this kernel to explore the state-space at a cost that is competitive with existing methods is rather important. As several articles have considered the behaviour of MCMC in high-dimensions (e.g.~\cite{high_dim}) and sometimes its success, one might expect that this aspect is not a bottleneck to its application. None-the-less, many high-dimensional MCMC kernels need care in their design, so the application of SMCMC is far from fool-proof. Intrinsic to the application of the method for filtering is an approximation of the filtering distribution: the mathematical performance of the algorithm, taking into account this approximation and when considering high-dimensions is yet to be studied and an important line of future work.}

\subsubsection*{Acknowledgements}
The work of HR and AJ was supported by KAUST baseline. The work of DC has been partially supported by European Research Council (ERC) Synergy grant STUOD-DLV-856408. NK was partially supported by a JP Morgan Chase AI Faculty award.

\subsection*{Data Availability Statement}

The data used in this study for the SWE initial values and boundary conditions are openly available from Copernicus Marine Service: Global Ocean Physics Analysis and Forecast at \url{https://doi.org/10.48670/moi-00016}. The drifters' data used in the SW model are openly available from NOAA at \url{https://doi.org/10.25921/x46c-3620}.

The code used to generate the numerical results presented in this study is available on Github at \url{https://github.com/ruzayqat/filtering_unknown_data_locations}.

\appendix

\begin{algorithm}[h]
 \textbf{Input:} The initial state $\breve Z_0=z_0$, the observations $\{Y_{t_k} = y_{t_k}\}_{k\geq 1}$, and the number of time discretization steps, $L$, between observations. Set $\tau_k=(t_k-t_{k-1})/L$. 

\begin{enumerate}
\item Initialize: For $l=0,\cdots,L-1$ compute $\breve Z_{(l+1)\tau_1} = \Phi(\breve Z_{l\tau_1},l\tau_1,(l+1)\tau_1)$. Compute $\tilde{Z}_{t_1}:=\breve{Z}_{t_1}+ W_{t_1}$, where $W_{t_1}\sim \mathcal{N}_d(0,\textcolor{black}{Q})$. \textcolor{black}{Run a standard RWM initialised at $\tilde{Z}_{t_1}$} to generate $N$ samples $\{Z_{t_1}^{(i)}\}_{i=1}^N$ from $\pi_1$ in \eqref{eq:pi_1_known}. Set $\widehat{\pi}_{1}^N(\varphi) \leftarrow \frac{1}{N} \sum_{i=1}^N \varphi(z_{t_1}^{(i)})$ and $k=2$.

\item For $k=2,\ldots,n$: 
\begin{enumerate}
\item For $i=1,\cdots, N$: compute 
$$\breve Z_{(l+1)\tau_k+t_{k-1}}^{(i)} = \Phi(\breve Z_{l\tau_k+t_{k-1}}^{(i)},l\tau_k+t_{k-1},(l+1)\tau_k+t_{k-1}),$$ 
where $0\leq l \leq L-1$ and $\breve Z^{(i)}_{t_{k-1}}= Z^{(i)}_{t_{k-1}}$.  
%\item Compute $\tilde{Z}^{(i)}_{t_k}:=\breve {Z}^{(i)}_{t_k}+ W^{(i)}_{t_k}$, where $W^{(i)}_{t_k}\sim \mathcal{N}_d(0,\textcolor{black}{Q})$. 
\item \textcolor{black}{Run Algorithm \ref{alg:pseudocode_alg2} to return samples $\{Z_{t_k}^{(i)}\}_{i=1}^N$}. Set $\widehat{\pi}_{k}^N(\varphi) \leftarrow \frac{1}{N} \sum_{i=1}^N \varphi(Z_{t_k}^{(i)})$. %Set $k \longleftarrow k+1$. %If $k=n+1$ go to the next step otherwise return to the start of step 2..
\end{enumerate}
\end{enumerate}
\textbf{Output:} Return $\{\widehat{\pi}_k^N(\varphi)\}_{k\in\{1,\cdots,n\}}$.
\caption{Pseudocode for Sequential MCMC Method for Filtering for $n$ time steps.}
\label{alg:pseudocode_alg1}
\end{algorithm}

\begin{algorithm}[h]
% \textbf{MCMC Step}
%Put $\mathsf{P}=\emptyset$, the empty set. For $i \in \{1\cdots,N+N_{\text{burn}}\}$, do the following:
\textcolor{black}{
Initialization: 
\begin{itemize}
\item Sample the auxiliary variable $j_0 \sim p(j)$ (uniformly).
\item Retrieve $\breve Z_{t_k}^{(j_0)}$ from Algorithm \ref{alg:pseudocode_alg1} and set $Z_{t_k}^{(0)} \longleftarrow\breve {Z}^{(j_0)}_{t_k}+ W_{t_k}$, where $W_{t_k}\sim \mathcal{N}_d(0,\textcolor{black}{Q})$. 
\item Compute
$
\pi_{\text{old}} \longleftarrow  g_k(Z_{t_k}^{(0)},y_{t_k}) ~f_k(Z_{t_{k-1}}^{(j_0)},Z_{t_k}^{(0)}).$ 
\end{itemize}
For $i ={1,\ldots,N+N_{\text{burn}}}$:
\begin{enumerate}
%
%Compute 
%$
%\pi_{\text{old}} \longleftarrow  g_k(Z_{t_k}',y_{t_k}) ~f_k(Z_{t_{k-1}}^{(j_i)},Z_{t_k}').$ 
%
%\item Compute 
%$
%\pi_{\text{old}} \longleftarrow  g_k(Z_{t_k}',y_{t_k}) ~f_k(Z_{t_{k-1}}^{(j_i)},Z_{t_k}').$  Note that in the evaluation of $f_k$ one needs $Z_{t_k}^{(j_i)}$.
\item Compute proposal for the:
\begin{itemize}
\item  state: $Z'_{t_k} \longleftarrow Z_{t_k}^{(i-1)} + W'$, where $W' \sim \mathcal{N}_d(0,{Q}')$; %\textcolor{black}{(${Q}'$ is proposal covariance matrix)}. 
\item auxiliary variable $j' = \left\{\begin{array}{ll}
		j_{i-1}-1 & \textrm{if}~j_{i-1}\notin\{1,N\}~\textrm{{w.p.}}~q\\
		j_{i-1} &  \textrm{if}~j_{i-1}\notin\{1,N\}~\textrm{{w.p.}}~1-2q\\
		j_{i-1}+1 & \textrm{if}~j_{i-1}\notin\{1,N\}~ \textrm{{w.p.}}~q\\
j_{i-1}+1 & \textrm{if}~j_{i-1}=1\\
j_{i-1}-1 & \textrm{if}~j_{i-1}=N
		\end{array}\right.$, where $q\in(0,\frac{1}{2}]$.
\end{itemize}
\item Retrieve $ Z_{t_{k-1}}^{(j')}$ from $\widehat\pi^N_{k-1}$ % and set $\tilde Z_{t_k} \longleftarrow Z_{t_k}^{(j_i)} $.
and compute 
$
\pi_{\text{new}} \longleftarrow  \left\{\begin{array}{ll}
		g_k(Z_{t_k}',y_{t_k}) ~f_k(Z_{t_{k-1}}^{(j')},Z_{t_k}') & \textrm{if}~j_{i-1}\notin\{1,N\}\\
		g_k(Z_{t_k}',y_{t_k}) ~f_k(Z_{t_{k-1}}^{(j')},Z_{t_k}') q &  \textrm{if}~j_{i-1}\in\{1,N\}
		\end{array}\right.  .$ %Note that in the evaluation of $f_k$ one needs $Z_{t_k}^{(j_i)}$. 
\item  Compute $\alpha = \min\{1, \pi_{\text{new}}/\pi_{\text{old}}\}$. Sample $u \sim \mathcal{U}[0,1]$. \begin{itemize}
\item  If $ u<\alpha$ set $Z_{t_k}^{(i)}\longleftarrow Z_{t_k}'$, $j_i=j'$ and $\pi_{\text{old}}\longleftarrow\pi_{\text{new}}$.
\item Else set $Z_{t_k}^{(i)}\longleftarrow Z_{t_k}^{(i-1)}$ and $j_{i}=j_{i-1}.$
\end{itemize}
\end{enumerate}
Return the sequence of $N$ final samples $\{Z_{t_k}^{(i)}\}_{i=N_{\text{burn}}+1}^N$.
}
\caption{Pseudocode for RWM to sample from $\pi_k^{N}(z_{t_k},j)$}
\label{alg:pseudocode_alg2}
\end{algorithm}

\section{Pseudocode for Algorithm \ref{alg:1} with \textcolor{black}{Reduced} Cost}\label{sec:pseudocode}
Consider the model in Example \ref{exam:pde} in addition to the observational model given by
\begin{equation*}
Y_{t_k} = \mathscr{O}_{t_k}(Z_{t_k}) + V_{t_k}, \quad V_{t_k} \stackrel{\textrm{i.i.d.}}{\sim} \mathcal{N}_{d_y}(0,\sigma_y^2 I_{d_y}),\quad t_k \in \mathsf{T},
\end{equation*}
where $\mathscr{O}_{t_k}:\mathsf{Z}\to \mathsf{Y}^{N_d}$ is an $\mathbb{R}^{d_y}$-vector valued function. We give below in Algorithms \ref{alg:pseudocode_alg1}-\ref{alg:pseudocode_alg2} the pseudocode to implement Algorithm \ref{alg:1} for this example with a cost of $\mathcal{O}(dN)$. Using an auxiliary variable method (e.g. see \cite[Section 5]{besag93})  \textcolor{black}{instead of sampling from 
$$
\pi_k^N(z_{t_k}) \propto g_k(z_{t_k},{y}_{t_k}) \frac{1}{N} \sum_{i=1}^Nf_k(z_{t_{k-1}}^{(i)},z_{t_k}),
$$
one samples from the joint distribution
\begin{align*}
\pi_k^{N}(z_{t_k},j) \propto g_k(z_{t_k},{y}_{t_k}) f_k(z_{t_{k-1}}^{(j)},z_{t_k})p(j)
\end{align*}
where $p(j)$ is uniform distribution over $\{1,\cdots, N\}$ so that $\pi_k^{N}(z_{t_k},j)$ admits $\pi_k^N(z_{t_k})$ as its marginal.} 
In practice, one would run $M$ independent runs of Algorithms \ref{alg:pseudocode_alg1}-\ref{alg:pseudocode_alg2} in parallel then use averages for performing inference.

Algorithm~\ref{alg:pseudocode_alg2} discards the first $N_{\text{burn}}$ steps required to converge to stationarity. In addition, note that the specification of $Q'$ in Algorithm~\ref{alg:pseudocode_alg2} (the covariance of the proposal distribution) requires careful design. When considering the SW model
%, for example, one must pay close attention to the selection of $W'$ in the MCMC step above. 
our approach was to construct a noise similar to the one described in Section \ref{subsec:rSWE_known}. In our simulation we set $W'$ as $W_{t_k}$ in Section \ref{subsec:rSWE_known} except that the random Gaussians $\epsilon_{t_k}^{\cdot,(i,j)}$ are sampled from $\mathcal{N}(0,\sigma'^2/(i\vee j+1))$ where \textcolor{black}{$\sigma' = 5\times 10^{-2}$}. This choice leads to an acceptance ratio $\alpha$ of the range of 0.2--0.3. For the auxiliary variable we used a simple random walk with $q=0.33$. This is a simple option that was effective here, but more elaborate schemes are possible by defining appropriate discrete transition matrices for $\{1,\ldots,N\}$.

\textcolor{black}{We also note that the implementation presented in Algorithms~\ref{alg:pseudocode_alg1}-\ref{alg:pseudocode_alg2} is not the most efficient and significant computational savings are possible. The current presentation was chosen for the sake of clarity, but it should be noted that step 2 (a) in Algorithm~\ref{alg:pseudocode_alg1} does not need to be executed for every $i\in\{1,\ldots,N\}$ but only for the sampled indices $\{j_i\}_{i=1}^{N_{\text{burn}}+N}$. This means that one can move step 2(a) of Algorithm~\ref{alg:pseudocode_alg1} in Algorithm~\ref{alg:pseudocode_alg2} and compute $Z_{t_k}^{(i)}$ only once (and then save in memory) for each $i\in \{j_i\}_{i=1}^{N_{\text{burn}}+N}$. Given this involves an expensive PDE evolution between $Z_{t_{k-1}}^{(i)}$ and $Z_{t_k}^{(i)}$ and the number of unique elements in $\{j_i\}_{i=1}^{N_{\text{burn}}+N}$ can be considerably less than $N$ (due to the rejections and random walk exploration), this approach can lead to a substantial computational saving.}

\section{Numerical solution of the SWE} \label{sec:num_pde_solver}

To write the SW equations in a compact form, we introduce the following vectors

\begin{align*}
U=[\eta,\eta u,\eta v]^{\top},\quad A(U) = [\eta u,\eta u^2 +\tfrac{1}{2}g \eta^2,\eta uv]^{\top}, \quad B(U) = [\eta v,\eta uv,\eta v^2 +\tfrac{1}{2}g\eta^2]^{\top},\\[0.2cm]
 C(U) = \Big[0,g\eta\frac{\partial H}{\partial x},g\eta\frac{\partial H}{\partial y}\Big]^{\top},  \quad D(U)=[0,f\eta v,-f\eta u]^{\top}. \qquad
\end{align*}
As a result, we can write the SWEs as
\begin{equation*}
U_t + A(U)_x + B(U)_y = C(U) + D(U).
\end{equation*}
Here $A$ and $B$ are the physical fluxes in the $x$ and $y$ directions, respectively. 

The spatial resolutions in the $x$ and $y$ directions are obtained via $\Delta_x  = (\underline{L}_x-\bar{L}_x)/N_x$ and $\Delta_y=(\underline{L}_y-\bar{L}_y)/N_y$. We  refer to the grid 
\begin{equation*}
\big\{(x_i,y_j) \in [\underline{L}_x,\bar{L}_x]\times[\underline{L}_y,\bar{L}_y]:\, x_i = (i-1)\Delta_x, \, y_j = (j-1) \Delta_y, \, ~ i \in \{1,\ldots,N_x\}, ~ j \in \{1,\ldots,N_y\}\big\}
\end{equation*}
as the physical grid.

Consider the uniform grid with finite volume cells $I_{i,j}=[x_{i-1/2},x_{i+1/2}] \times [y_{j-1/2},y_{j+1/2}]$ centered at $(x_i,y_j)=(\frac{x_{i-1/2}+x_{i+1/2}}{2},\frac{y_{j-1/2}+y_{j+1/2}}{2})$, for all $(i,j) \in \{0,\ldots, N_x+1\}\times  \{0,\ldots, N_y+1\}$, with grid size of $(N_x+2)\times(N_y+2)$. Then, the discretized solution of the SWEs on $[t_{k-1}, t_k)$ is as follows. For $l \in \{0,\cdots,L-2\}$, we compute
\begin{equation*}
U_{i,j}^{t_{k-1}+(l+1)\tau_k} = U_{i,j}^{t_{k-1}+l\tau_k} - \frac{\tau_k}{\Delta_x} (A_{i+\frac{1}{2},j}^{*} - A_{i -\frac{1}{2},j}^{*}) - \frac{\tau_k}{\Delta_y} (B_{i,j+\frac{1}{2}}^{*} - B_{i ,j-\frac{1}{2}}^{*}) + \tau_k C_{i,j}^{t_{k-1}+l\tau_k} +\tau_k D_{i,j}^{t_{k-1}+l\tau_k},
\end{equation*}
where $A^{*}$ and $B^{*}$ are the numerical Lax-Friedrichs fluxes given by
\begin{align*}
A_{i+\frac{1}{2},j}^{*} &= \tfrac{1}{2} [A(U_{i,j}^{t_{k-1}+l\tau_k})+A(U_{i+1,j}^{t_{k-1}+l\tau_k})]  - \tfrac{1}{2} \lambda_{i+\frac{1}{2},j,\max}^x [U_{i+1,j}^{t_{k-1}+l\tau_k} - U_{i,j}^{t_{k-1}+l\tau_k}]\\
A_{i-\frac{1}{2},j}^{*} &= \tfrac{1}{2} [A(U_{i,j}^{t_{k-1}+l\tau_k})+A(U_{i-1,j}^{t_{k-1}+l\tau_k})]  - \tfrac{1}{2} \lambda_{i-\frac{1}{2},j,\max}^x [U_{i,j}^{t_{k-1}+l\tau_k} - U_{i-1,j}^{t_{k-1}+l\tau_k}]\\
B_{i,j+\frac{1}{2}}^{*} &= \tfrac{1}{2} [B(U_{i,j}^{t_{k-1}+l\tau_k})+B(U_{i,j+1}^{t_{k-1}+l\tau_k})]  - \tfrac{1}{2} \lambda_{i,j+\frac{1}{2},\max}^y [U_{i,j+1}^{t_{k-1}+l\tau_k} - U_{i,j}^{t_{k-1}+l\tau_k}]\\
B_{i,j-\frac{1}{2}}^{*} &= \tfrac{1}{2} [B(U_{i,j}^{t_{k-1}+l\tau_k})+B(U_{i,j-1}^{t_{k-1}+l\tau_k})]  - \tfrac{1}{2} \lambda_{i,j-\frac{1}{2},\max}^y [U_{i,j}^{t_{k-1}+l\tau_k} - U_{i,j-1}^{t_{k-1}+l\tau_k}],
\end{align*}
therefore, we have
\begin{align*}
A_{i+\frac{1}{2},j}^{*} - A_{i-\frac{1}{2},j}^{*} &= \tfrac{1}{2} [A(U_{i+1,j}^{t_{k-1}+l\tau_k})-A(U_{i-1,j}^{t_{k-1}+l\tau_k})]  \\  & \qquad\qquad - \tfrac{1}{2} \Big(\lambda_{i+\frac{1}{2},j,\max}^x [U_{i+1,j}^{t_{k-1}+l\tau_k} - U_{i,j}^{t_{k-1}+l\tau_k}] - 
 \lambda_{i-\frac{1}{2},j,\max}^x [U_{i,j}^{t_{k-1}+l\tau_k} - U_{i-1,j}^{t_{k-1}+l\tau_k}] \Big)\\[0.3cm]
B_{i,j+\frac{1}{2}}^{*} - B_{i,j-\frac{1}{2}}^{*} &= \tfrac{1}{2} [B(U_{i,j+1}^{t_{k-1}+l\tau_k})-B(U_{i,j-1}^{t_{k-1}+l\tau_k})] \\& \qquad\qquad - \tfrac{1}{2} \Big( \lambda_{i,j+\frac{1}{2},\max}^y [U_{i,j+1}^{t_{k-1}+l\tau_k} - U_{i,j}^{t_{k-1}+l\tau_k}] -
 \lambda_{i,j-\frac{1}{2},\max}^y [U_{i,j}^{t_{k-1}+l\tau_k} - U_{i,j-1}^{t_{k-1}+l\tau_k}]\Big),
\end{align*}
where $\lambda_{i^*,j^*,\max}^x$ is the maximum eigenvalue of the Jacobian matrix $\partial A(U)/\partial U$ evaluated at $U_{i^*,j^*}^{t_{k-1}+l\tau_k}$. The eigenvalues are $\{u_{i^*,j^*}\pm \sqrt{g\eta_{i^*,j^*}},u_{i^*,j^*}\}$. We set $\lambda_{i^*,j^*,\max}^x =|u_{i^*,j^*}|+\sqrt{g\eta_{i^*,j^*}}$. Similarly, $\lambda_{i^*,j^*,\max}^y$ is the maximum eigenvalue of the Jacobian matrix $\partial B(U)/\partial U$ evaluated at $U_{i^*,j^*}^{t_{k-1}+l\tau_k}$ and we take it to be $|v_{i^*,j^*}|+\sqrt{g\eta_{i^*,j^*}}$. Then the hidden signal at time $ t= t_{k-1} + l\tau_k\in [t_{k-1}, t_k)$, is the vector given by
\begin{equation*}
Z_t = [(\eta_{i}^t)_{1\leq i \leq N_xN_y }, (u_{i}^t)_{1\leq i \leq N_xN_y },(v_{i}^t)_{1\leq i \leq N_xN_y } ]^\top \in \mathbb{R}^{3N_xN_y},
\end{equation*}
where $Z_0$ is known. Here the vectors $(\eta_{i}^t)_{1\leq i \leq N_xN_y }, (u_{i}^t)_{1\leq i \leq N_xN_y },(v_{i}^t)_{1\leq i \leq N_xN_y }$ are obtained from the approximate solution $U_{i,j}^t$, $(i,j)\in \{0,\cdots,N_x+1\} \times \{0,\cdots,N_y+1\}$.

\section{Implementation of the noise } \label{appx:noise}
\textcolor{black}{
\begin{figure}[h!]
\centering
\includegraphics[scale=0.22]{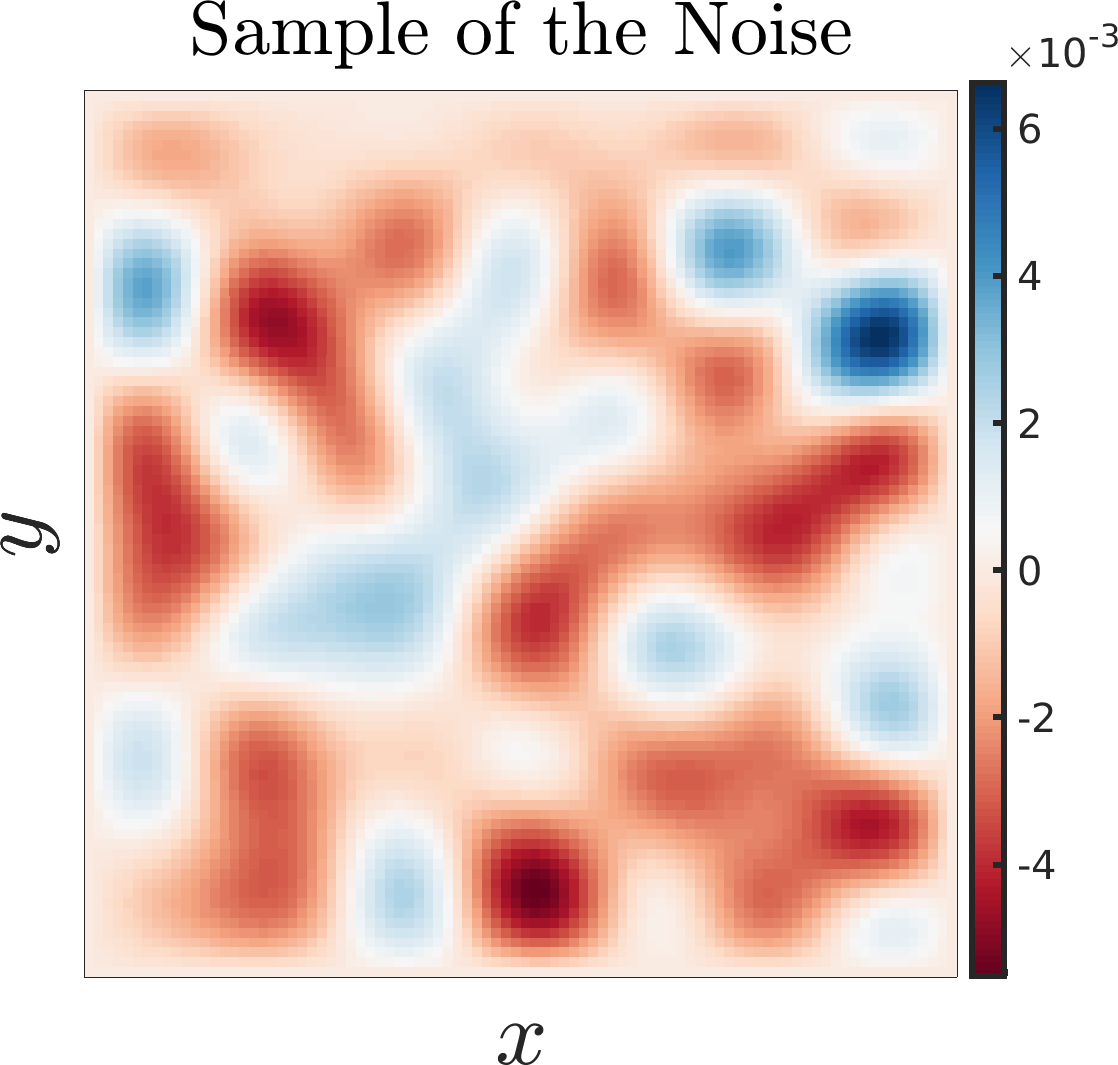}
\caption{\textcolor{black}{A random sample of the noise as described in this section.}}
\label{fig:noise_sample}
\end{figure}
We show how the noise is  computed for each of the $\eta,u,v$ variables. As the procedure is the same in each case, we drop the $\eta,u,v$ superscripts, and define
\begin{equation*}
S_1 = \left[\sin\left(\tfrac{\pi j y_l}{\bar{L}_y-\underline{L}_y}\right)\right]_{l=1, j = 0}^{N_y, J-1} \in \mathbb{R}^{N_y \times J}, \qquad S_2 = \left[\sin\left(\tfrac{\pi i x_s}{\bar{L}_x-\underline{L}_x}\right)\right]_{s=1, i = 0}^{N_x, J-1} \in \mathbb{R}^{N_x \times J},
\end{equation*}
and let ${\epsilon}_{t_k}$ be a random $J\times J$ matrix with independent entries, $\epsilon_{t_k}^{ij} \sim \mathcal{N}(0,\sigma^2/(i\vee j+1))$, for $i,j \in \{0,\cdots,J-1\}$, with $\sigma > 0$. Define $\Xi_{t_k} := S_1 {\epsilon}_{t_k} S_2^T \in \mathbb{R}^{N_y\times N_x}$ and let  $X_{t_k}:=\text{Vec}(\Xi_{t_k}) \in \mathbb{R}^{N_xN_y}$. Then $X_{t_k}$ is a random sample of the noise, and one has ${\mathbb{E}}[X_{t_k}] = 0$ %(the expectation is taken with respect to the random matrix $\epsilon_{t_k}$) 
and
\begin{equation*}
\mathbb{C}ov[X_{t_k}] = {\mathbb{E}}[X_{t_k}X_{t_k}^T].
\end{equation*}
We drop the subscript $t_k$ from $\Xi_{t_k}$, then
{\small
\begin{align*}
&X_{t_k}X_{t_k}^T=\\
& \left[\begin{array}{ccc|ccc|cc|ccc}
\Xi_{11}^2 & \cdots & \Xi_{11}\Xi_{N_y1} & \Xi_{11}\Xi_{12} & \cdots & \Xi_{11}\Xi_{N_y2} & \cdots & \cdots & \Xi_{11}\Xi_{1N_x} & \cdots & \Xi_{11}\Xi_{N_yN_x} \\
\vdots & \ddots & \vdots & \vdots & \ddots & \vdots & \cdots & \cdots & \vdots &\ddots & \vdots \\
\Xi_{N_y1}\Xi_{11} & \cdots & \Xi_{N_y1}^2 & \Xi_{N_y1}\Xi_{12} & \cdots &  \Xi_{N_y1}\Xi_{N_y2} & \cdots & \cdots & \Xi_{N_y1}\Xi_{1N_x} & \cdots & \Xi_{N_y1}\Xi_{N_yN_x} \\
\hline
\Xi_{12}\Xi_{11} & \cdots & \Xi_{12}\Xi_{N_y1} & \Xi_{12}^2 & \cdots & \Xi_{12}\Xi_{N_y2} & \cdots & \cdots & \Xi_{12}\Xi_{1N_x} & \cdots & \Xi_{12}\Xi_{N_yN_x} \\
\vdots & \ddots & \vdots & \vdots & \ddots & \vdots & \cdots & \cdots  & \vdots & \ddots & \vdots \\
\Xi_{N_y2}\Xi_{11} & \cdots & \Xi_{N_y2}\Xi_{N_y1} & \Xi_{N_y2}\Xi_{12} & \cdots & \Xi_{N_y2}^2 & \cdots & \cdots & \Xi_{N_y2}\Xi_{1N_x} & \cdots & \Xi_{N_y2}\Xi_{N_yN_x} \\
\hline
\vdots & \vdots & \vdots & \vdots & \vdots & \vdots & \ddots &  & \vdots & \vdots & \vdots \\
\vdots & \vdots & \vdots & \vdots & \vdots & \vdots &  & \ddots & \vdots & \vdots & \vdots \\
\hline
\Xi_{1N_x}\Xi_{11} & \cdots & \Xi_{1N_x}\Xi_{N_y1} & \Xi_{1N_x}\Xi_{12} & \cdots & \Xi_{1N_x}\Xi_{N_y2} & \cdots & \cdots & \Xi_{1N_x}^2 & \cdots & \Xi_{1N_x}\Xi_{N_yN_x} \\
\vdots & \ddots & \vdots & \vdots & \ddots & \vdots & \cdots & \cdots &  \vdots & \ddots & \vdots \\
\Xi_{N_yN_x}\Xi_{11} & \cdots & \Xi_{N_yN_x}\Xi_{N_y1} & \Xi_{N_yN_x}\Xi_{12} & \cdots & \Xi_{N_yN_x}\Xi_{N_y2} & \cdots & \cdots & \Xi_{N_yN_x}\Xi_{1N_x} & \vdots & \Xi_{N_yN_x}^2 
\end{array}
\right]  
\end{align*}
}
where $\Xi_{ij}$ is the element of $\Xi$ in the $i^{\text{th}}$ row and the $j^{\text{th}}$ column. The noise covariance matrix is then the above matrix with element-wise expectations given by 
\begin{align*}
&\mathbb{E}[\Xi_{i_1j_1}\Xi_{i_2j_2}]=\sigma^2\sum_{m,n=0}^{J-1} S_1^{i_1m}S_1^{i_2m}S_2^{j_1n}S_2^{j_2n} \frac{1}{m\vee n + 1}.
\end{align*}
In Figure \ref{fig:noise_sample}, we show a random sample of the noise used in the rotating SW model.
}

%%%%%%%%%%%%%%%%%%%%%

\end{document}